\documentclass[11pt]{article} \hoffset=-15mm \voffset=-10mm
\usepackage{graphicx,amsmath,amssymb,epsf} \usepackage{latexsym,bm,
  slashed} \usepackage{xcolor} \definecolor{dark}{rgb}{0.10,0.2,0.3}
\definecolor{magenta}{rgb}{0.7,0.1,0.3}
\definecolor{purpure}{rgb}{0.5,0.15,0.3}
\usepackage[font=small,format=plain,labelfont=bf,up,textfont=it,up]{caption}
\usepackage{hyperref, cite} \hypersetup{colorlinks, citecolor=blue,
  filecolor=blue, linkcolor=magenta,
  urlcolor=purpure,hyperfootnotes=true,pdftex} 

\newcommand{\tr}{{\rm tr}}

 \title{\bf  \Large 
The Color Glass Condensate formalism, Balitsky-JIMWLK evolution  and  Lipatov's high energy   effective action     
   } \author{{\Large Martin Hentschinski} 
\bigskip \\  Departamento de Actuaria, F\'isica y Matem\'aticas \\
Universidad de las Am\'ericas Puebla \\
Santa Catarina Martir, 72820 \\
Puebla, Mexico
}

\begin{document}

\maketitle
\begin{abstract}
  We investigate the question whether Lipatov's high energy effective
  action is capable to reproduce quark and gluon propagators which
  resum interaction with a strong background field within high energy
  factorization. Such propagators are frequently employed in
  calculations within the Color Glass Condensate formalism, in
  particular when considering scattering of a dilute projectile on a
  dense target nucleus or nucleon. We find that such propagators can
  be obtained from the high energy effective action, if a special
  parametrization of the gluonic field is used, first proposed by
  Lipatov in the original publication on the high energy effective
  action. The obtained propagators are  used to rederive from the
  high energy effective action the leading order Balitsky-JIMWLK
  evolution equation in covariant gauge.  As an aside, our result
  confirms the definition of the reggeized gluon as the logarithm of
  an adjoint Wilson lines, proposed in the literature.
\end{abstract}

\section{Introduction}
\label{sec:introduction}

The Color Glass Condensate (CGC) formalism is an effective field
theory approach to Quantum Chromodynamics (QCD) at small $x$ where
gluon densities in the nucleus or proton are large.  With $x$ the
ratio of the hard scale $M^2$ of a certain hard process and $s$ the
center-of-mass energy squared, the limit $x \to 0$ at fixed $M^2$
corresponds to the perturbative Regge limit of QCD. In such a scenario, the
smallness of the strong coupling $\alpha_s(M^2) \ll 1$ can be
compensated by logarithms in $x$, $\alpha_s(M^2) \ln 1/x \sim 1$,
which requires the resummation of terms
$\left( \alpha_s(M^2) \ln 1/x \right)^n$ to all orders. For
perturbative scattering amplitudes, such a resummation is achieved at
leading
\cite{Fadin:1975cb,Balitsky:1978ic} and
next-to-leading order \cite{Fadin:1998py} by the
Balitsky-Fadin-Kuraev-Lipatov (BFKL) evolution equation. Even though
BFKL evolution is successfully applied to the description of collider
data at currently accessible center-of-mass energies, see {\it e.g.}
\cite{Bautista:2016xnp}, the power-like rise of the gluon
distribution predicted by BFKL evolution will eventually drive
cross-sections to a region of phase space, where parton densities are
no longer perturbative; BFKL evolution will therefore break down in
such a regime.  Instead it is more appropriate to treat the
hadron or nucleus as a coherent color field rather than a collection
of incoherent and individual partons. This is  the region of phase
space which is addressed by the initially mentioned CGC, see~\cite{gijv}
for a review.  At the classical level, the CGC generalizes scattering
via exchange of a single gluon to multiple gluon exchanges within high
energy factorization. Including furthermore quantum effects, one
arrives at a resummation of logarithms in $1/x$, generalizing BFKL
evolution to the case of large gluon densities. The resulting
Balitsky-JIMWLK evolution~\cite{Balitsky:1995ub,
  jimwlk1,jimwlk2,jimwlk6,jimwlk8} provides finally an
evolution equation for Wilson lines which sum up the strong gluonic
field in the
target. \\

In the present article we discuss Lipatov's high energy effective
action~\cite{Lipatov:1995pn,Lipatov:1996ts} and its relation to the
above mentioned formulation of an CGC effective theory. One of the
main advantages of Lipatov's high energy effective action is that it
provides a gauge invariant factorization of QCD amplitudes in the high
energy limit through introducing a new type of field, {\it i.e.} the
reggeized gluon. Using this effective action it has been possible to
both reproduce and derive a number of next-to-leading order (NLO)
results, most notable the calculation of NLO correction to forward jet
production without \cite{quarkjet,gluonjet} and with rapidity
gap~\cite{Hentschinski:2014esa}, the gluon Regge trajectory up to two
loop~\cite{traject}, and the NLO kernel of the
Bartels-Kwiecinski-Praszalowicz evolution equation
\cite{Bartels:2012sw}, see also the review \cite{review}; for the
determination of NLO corrections for reggeized quarks see
\cite{Nefedov:2017qzc}.  The description of scattering amplitudes for
multiple reggeized gluon exchange has been also studied by a number of
authors, see {\it e.g.}
\cite{Braun:2017qij,Hentschinski:2009zz,Hentschinski:2009ga}.  At the
same time the ability of Balitsky-JIMWLK evolution to reproduce
scattering amplitudes with multiple reggeized gluon states has been
demonstrated for various cases, see {\it e.g.}
\cite{Bartels:2004ef,Ayala:2014nza}, hinting at a possible equivalence
of both formalisms.  Furthermore the Color Glass Condensate formalism
and the high energy effective action have been compared directly at the
level of the Lagrangian, see {\it e.g.}  \cite{jimwlk2, Hatta:2005rn,
  Bondarenko:2018pvv}. In particular \cite{ Bondarenko:2018pvv}
demonstrate that it is possible to reproduce the classical gluon
fields of the CGC approach from the Lipatov's high energy effective
action.
\\

Instead of comparing the two approaches on the level of the resulting
effective Lagrangians, we take here a pragmatic approach and attempt
to answer the question whether Lipatov's high energy effective action
can be used to reproduce the quark and gluon propagators in the
presence of a strong gluonic field. Such propagators are one of the core
elements in calculations of scattering of dilute projectiles on dense
targets within the Color-Glass-Condensate formalism. We find that this
can be indeed achieved by choosing a special parametrization of the
gluonic field already proposed in~\cite{Lipatov:1995pn}. Moreover,
since Lipatov's high energy effective action provides a gauge
invariant factorization of QCD amplitudes in the high energy limit,
the resulting propagators are not restricted to a certain gauge, such
as light-cone gauge. The obtained propagators allow furthermore to rederive
leading order Balitsky-JIMWLK evolution directly from Lipatov's high
energy effective action.  As an aside,
our result confirms that the definition of the reggeized gluon as the
logarithm of an adjoint Wilson lines, proposed in \cite{Caron-Huot:2013fea}, is
consistent with Lipatov's high energy effective action.
\\

The outline of this paper is as follows. Sec.~\ref{sec:eff} provides a
short summary of Lipatov's high energy effective
action. Sec.~\ref{sec:parametrize} introduces the special
parametrization of the gluonic field proposed in~\cite{Lipatov:1995pn}
and demonstrates how it can be used to derive resummed partonic
propagators in the presence of a strong reggeized gluon
field. Sec.~\ref{sec:literature} contains a comparison of our result
with the literature.  Sec.~\ref{sec:balitsky-jimwlk-evol} presents a
derivation of Balitsky-JIMWLK evolution from Lipatov's high energy
effective action. In Sec.~\ref{sec:conclusion-summary} we summarize
our results and draw our conclusions. Some details of the calculations
are summarized in two appendices.

\section{The High-Energy Effective Action}
\label{sec:eff}

\begin{figure}[th]
       \centering
   \parbox{.7cm}{\includegraphics[height = 1.8cm]{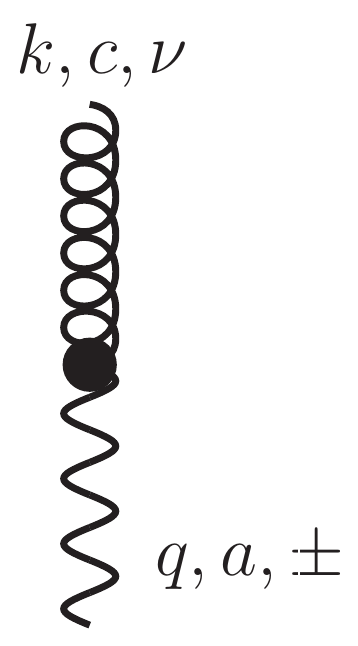}} $=  \displaystyle 
   \begin{array}[h]{ll}
    \\  \\ \frac{- i}{2}{\bm q}^2 \delta^{a c} (n^\pm)^\nu,  \\ \\  \qquad   k^\pm = 0.
   \end{array}  $ 
 \parbox{1.2cm}{ \includegraphics[height = 1.8cm]{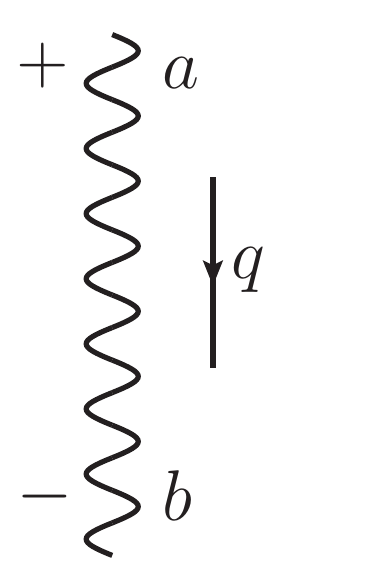}}  $=  \displaystyle    \begin{array}[h]{ll}
    \delta^{ab} \frac{ 2 i}{{\bm q}^2} \end{array}$,
 \parbox{1.7cm}{\includegraphics[height = 1.8cm]{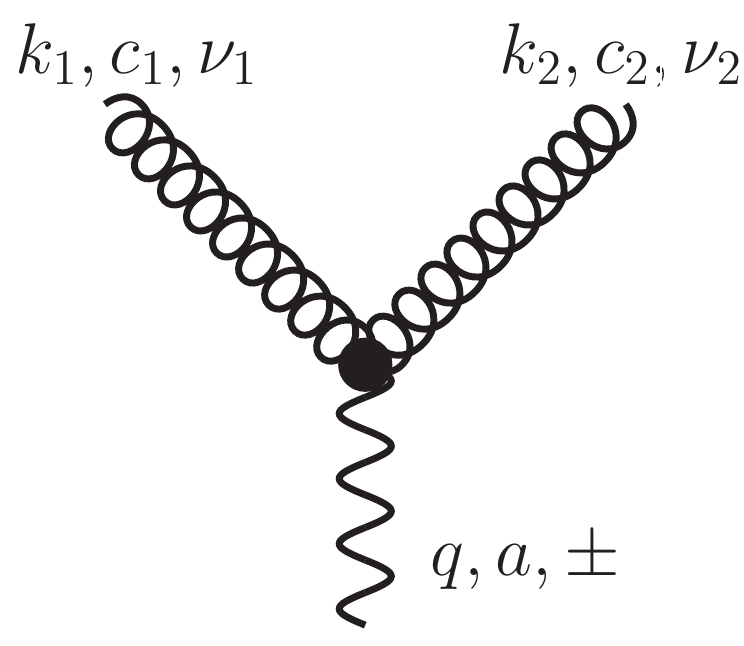}} $ \displaystyle  =  \begin{array}[h]{ll}  \\ \\ \frac{g}{2} f^{c_1 c_2 a} \frac{{\bm q}^2}{k_1^\pm}   (n^\pm)^{\nu_1} (n^\pm)^{\nu_2},  \\ \\ \quad  k_1^\pm  + k_2^\pm  = 0,
 \end{array}$
 \\
\parbox{3cm}{\center (a)} \parbox{4cm}{\center (b)} \parbox{5cm}{\center (c)}

\vspace{1cm}
  \parbox{2.4cm}{\includegraphics[height = 1.8cm]{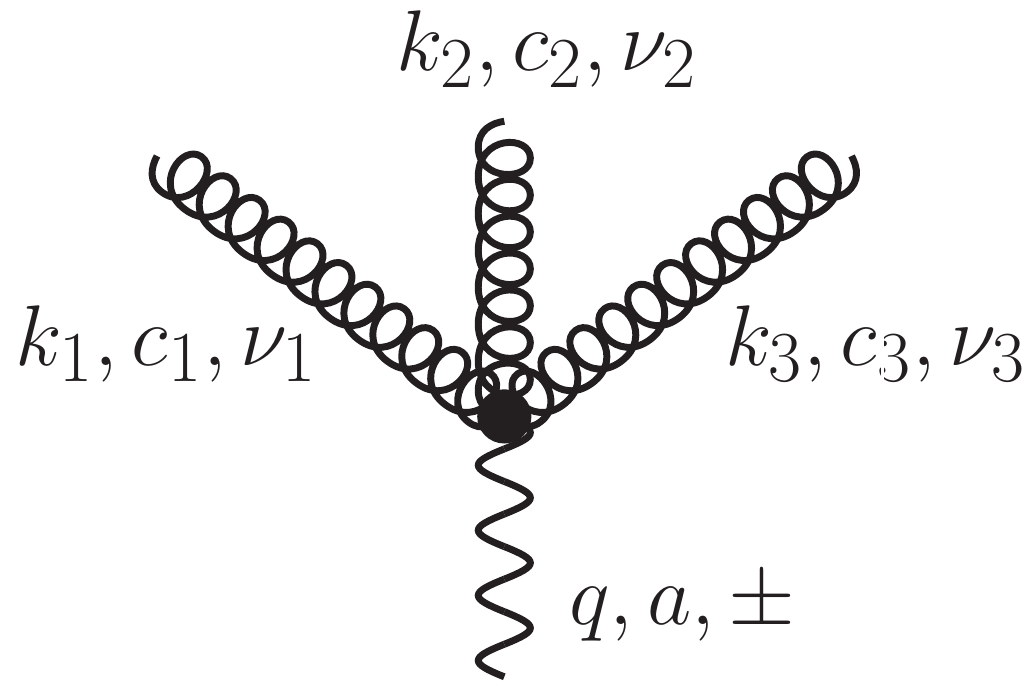}} $ \displaystyle 
   \begin{array}[h]{l}  \displaystyle  \\ \displaystyle= \frac{ig^2}{2} {\bm{q}}^2 
\left(\frac{f^{a_3a_2 e} f^{a_1ea}}{k_3^\pm k_1^\pm} 
+
 \frac{f^{a_3a_1 e} f^{a_2ea}}{k_3^\pm k_2^\pm}\right) (n^\pm)^{\nu_1} (n^\pm)^{\nu_2} (n^\pm)^{\nu_3}, \\ \\
\qquad \qquad   k_1^\pm + k_2^\pm + k_3^\pm = 0.
   \end{array}
$ \\ 
\vspace{.3cm}
\parbox{1cm}{(d)}
\caption{\small Feynman rules for the lowest-order effective vertices
  of the effective action. Wavy lines denote reggeized fields and
  curly lines gluons. Note that in comparison with the Feynman rules
  used in \cite{quarkjet,gluonjet,Hentschinski:2014esa,traject} we
  absorbe a factor $1/2$ into the vertices which is compensated by
  changing the residue of the reggeized gluon propagator from $1/2$ to
  $2$.}
\label{fig:3}
\end{figure}

Within the framework provided by Lipatov's effective action
\cite{Lipatov:1995pn,Lipatov:1996ts}, QCD amplitudes are in the high
energy limit decomposed into gauge invariant sub-amplitudes which are
localized in rapidity space. The effective Lagrangian then describes
the coupling of quarks ($\psi$) and gluon ($v_\mu$) fields to a new
degree of freedom, the reggeized gluon field $A_\pm (x)$. The latter
is introduced as a convenient tool to reconstruct the complete QCD
amplitudes in the high energy limit out of the sub-amplitudes
restricted to small rapidity intervals.  Lipatov's effective action is
obtained by adding an induced term $ S_{\text{ind.}}$ to the QCD
action $S_{\text{QCD}}$,
\begin{align}
  \label{eq:effac}
S_{\text{eff}}& = S_{\text{QCD}} +
S_{\text{ind.}} ,
\end{align}
where the induced term $ S_{\text{ind.}}$ describes the coupling of
the gluonic field $v_\mu = -it^a v_\mu^a(x)$ to the reggeized gluon
field $A_\pm(x) = - i t^a A_\pm^a (x)$, with $t^a$  a
SU$(N_c)$ generator in the fundamental representation,
$\tr(t^at^b) = \delta^{ab}/2$. 
For the definition of light-cone directions we
follow the conventions established in the original publication
\cite{Lipatov:1995pn},
\begin{align}
   \label{eq:13}
   k^\pm & = n^\pm \cdot k = n_\mp \cdot k = k_\mp,
\end{align}
with $n^\pm  \cdot n^\mp = 2$ and $(n^\pm)^2 = 0$.  This implies the
following Sudakov decomposition of a four momentum
\begin{align}
  \label{eq:14}
  k & =  \frac{k^+}{2} n^- +   \frac{k^-}{2} n^+ + {\bm k} =  \frac{k_-}{2} n_+ +   \frac{k_+}{2} n_- + {\bm k}.
\end{align}
Note that transverse momenta and coordinates will be denoted by bold
letters.  Furthermore 
\begin{align}
  \label{eq:9}
  \partial_\pm x^\pm & = 2,   & \partial_\mp x^\pm & = 0\,.
\end{align}
High energy factorized amplitudes
reveal strong ordering in plus and minus components of momenta which
leads to the following kinematic constraint obeyed by the
reggeized gluon field:
\begin{align}
  \label{eq:kinematic}
  \partial_+ A_- (x)& = 0 = \partial_- A_+(x).
\end{align}
Even though the reggeized gluon field is charged under the QCD gauge
group SU$(N_c)$, it is defined to be invariant under local gauge
transformation $\delta_L A_\pm = 0$.  With the local gauge
transformations of gluon and quark fields given by
\begin{align}
  \label{eq:gauge_gluon}
\delta_{\text{L}} v\mu &= \frac{1}{g}[D_\mu, \chi_L], & \delta_{\text{L}} \psi &= -\chi_L \psi.
&
D_\mu & = \partial_\mu + g v_\mu,
\end{align}
where $D_\mu$  denotes the covariant derivative and $\chi_L$ the
parameter of the local gauge transformations which decreases for $x
\to \infty$, the reggeized gluons fields are {\it invariant} under 
 local gauge transformations,
\begin{align}
  \label{eq:localgauge}
     \delta_\text{L} A_\pm = \frac{1}{g}[A_\pm,
  \chi_L] = 0 \, .
\end{align}
The kinetic term and the gauge invariant coupling of the reggeized
gluon field to the QCD gluon field are provided by the induced term
\begin{align}
\label{eq:1efflagrangian}
  S_{\text{ind.}} & = \int \text{d}^4 x \, \bigg\{ 
\text{tr}\left[\left(T_-[v(x)] - A_-(x) \right)\partial^2_\perp A_+(x)\right]
\notag \\
& \hspace{3cm}
+\text{tr}\left[\left(T_+[v(x)] - A_+(x) \right)\partial^2_\perp
  A_-(x)\right] \bigg\}.
\end{align}
The functionals $T_\pm[v] $ can be obtained from the following
operator definition
\begin{align}
\label{eq2:efflagrangian}
T_\pm[v] =
&
%
-\frac{1}{g}\partial_\pm  \frac{1}{1 + \frac{g}{\partial_\pm}v_\pm}
 =  v_\pm - g  v_\pm\frac{1}{\partial_\pm} v_\pm + g^2 v_\pm
\frac{1}{\partial_\pm} v_\pm\frac{1}{\partial_\pm} v_\pm - \ldots
\notag \\
\end{align}
where the integral operator is implied to act on a unit constant
matrix from the left. Boundary conditions of the $1/\partial_\pm$ are fixed through
\begin{align}
  \label{eq:6}
 \frac{1}{1 + \frac{g}{\partial_\pm}v_\pm} &  = 
 \mathcal{P}\exp\bigg(-\frac{g}{2} \int_{-\infty}^{x^\pm}dx'^\pm  v_\pm(x')\bigg)
\notag \\ 
&= 
1 -\frac{g}{2} \int_{-\infty}^{x^\pm}dx'^\pm  v_\pm(x') + \frac{g^2}{4} \int_{-\infty}^{x^\pm}   dx^{'\pm}   \int_{-\infty}^{x^{'\pm}} dx^{''\pm}   v_\pm(x') v_\pm(x'')
 + \ldots
\end{align}
Due to the induced term in
Eq.~(\ref{eq:effac}), the Feynman rules of the effective action
comprise, apart from the usual QCD Feynman rules, the propagator of
the reggeized gluon and an infinite number of so-called induced
vertices.  The leading order vertices and propagators  are
summarized in Fig.~\ref{fig:3}.
These induced vertices are special in the sense that they contain only
the anti-symmetric color-octet sector of the eikonal operator
Eq.~\eqref{eq2:efflagrangian}.

While the projection on the color octet sector arises automatically
from the induced term due to the combination with the reggeized gluon
field, the anti-symmetric color structure (written in terms of
SU$(N_c)$ structure constants only) requires in general use of a
corresponding projector, for an explicit construction see
\cite{Hentschinski:2011xg}. The original argument given by Lipatov for
this projection is based on the observation that in generalized
Multi-Regge Kinematics the values of the operator $\partial_\pm$
acting on a gluonic field is never zero for the vertices arising from
Eq.~\eqref{eq:1efflagrangian}, since the resulting light-cone momenta
are proportional to large center of mass energies of clusters of particles
significantly separated in rapidity. In particular
\begin{align}
  \label{eq:19}
  \frac{1}{\partial_\pm} \tilde{v}_\pm(p) & = \frac{i}{p_\pm} \tilde{v}_\pm(p)
\end{align}
with $p_\pm \neq 0$ where $\tilde{v}_\pm(p)$ denotes the Fourier
transform of the gluonic field $v(x)$; this is especially true for the
case of real particle production within the generalized Multi-Regge
Kinematics, which initiated the discussion of the formulation of the
high energy effective action in \cite{Lipatov:1995pn}. For a more
detailed discussion we refer to \cite{Antonov:2004hh}.  With
$p_\pm \neq 0$, anti-symmetric color structure as given in
Fig.~\ref{fig:3} arises automatically from the high energy effective
action, see also the discussion in \cite{Antonov:2004hh}. The
condition $p_\pm \neq 0$ is however at least at first violated in the
evaluation of loop integrals, where the $p_\pm$ are integrated over
all possible values.  The projection of
\cite{Hentschinski:2011xg} implies then the use of the boundary
conditions of Eq.~\eqref{eq:6}, with an additional projection for the
color structure of the vertices Fig.~\ref{fig:3} on the desired
anti-symmetric color octet sector. Corresponding symmetric
counter-parts are then taken into account by exchange of multiple
reggeized gluons and combination of multiple reggeized gluons and
induced vertices, see also the discussion in Appendix
\ref{sec:multi-gluon-exchange}. In the following we use always the
pole prescription for induced vertices proposed in
\cite{Hentschinski:2011xg}.

\section{Resummation of a strong  reggeized gluon field}

\label{sec:parametrize}

In the following we provide a formulation of the high energy effective
action which allows for a straightforward resummation of multiple
reggeized gluon exchange in the chase of quasi-elastic scattering,
which is the relevant case for describing scattering of a dilute
partonic projectile on a dense target nucleus or proton.

\subsection{A special parametrization of the gluonic field}
\label{sec:spec-param-gluon}

The bulk of calculations performed within the framework set by the
high energy effective action employs the vertex Fig.~\ref{fig:3}.a)
which provides a direct transition between a reggeized gluon field and
a conventional QCD gluon. As noted in~\cite{Lipatov:1995pn,
  Lipatov:1996ts}, it is possible to avoid the use of such a direct
transition vertex, if one performs a shift
$v_\pm \to V_\pm = v_\pm + A_\pm$ of the gluonic field in the
effective action\footnote{Such a shift has been used for instance in
  \cite{Braun:2017qij, Hentschinski:2009zz}}.  Such a shift has however the disadvantage
that the gluonic field $v_\pm$ transforms like a gauge field under
local gauge transformations while the reggeized gluon field is
invariant under such transformations. To avoid such differing
transformation properties, the following parametrization of the
gluonic field has been proposed in~\cite{Lipatov:1995pn}:
\begin{align}
  \label{eq:para0}
  V^\mu(x) & = v^\mu(x) + \frac{n_+^\mu}{2}  U[v_+(x)] A_-(x)U^{-1}[v_+(x)] 
+
 \frac{n_-^\mu}{2}  U[v_-(x)] A_+(x)U^{-1}[v_-(x)] 
\notag \\
&= 
 v^\mu(x) + \frac{n_+^\mu}{2}   B_-(x) + \frac{n_-^\mu}{2}   B_+(x)  \, ,
\end{align}
where 
\begin{align}
  \label{eq:1}
  B_\pm[v_\mp] = U[v_\mp] A_\pm U^{-1}[v_\mp] \,.
\end{align}
and (inverse) Wilson line operators are   defined as 
\begin{align}
  \label{eq:U}
  U[v_\pm] &=  \frac{1}{1 + \frac{g}{\partial_\pm} v_\pm}, &  
U^{-1}[v_\pm]
  & = 1 + \frac{g}{\partial_\pm} v_\pm \, .
\end{align}
Here the integral operators $U$ and $U^{-1}$ act on a unit constant
matrix from the left- and right-had sides, respectively.  For the
above composite field $B_\pm[v_\mp]$, one finds  the following gauge
transformation properties:
\begin{align}
  \label{eq:deltaterm}
  \delta_L B_\pm & = \delta_L U[v_\mp] A_\pm U^{-1}[v_\mp] + U[v_\mp] A_\pm\delta_L U^{-1}[v_\mp]  = \left[g B_\pm, \chi_L \right]\,.
\end{align}
As a consequence  the shifted gluonic field
Eq.~\eqref{eq:para0} transforms as
\begin{align}
  \label{eq:deltaterm2}
    \delta  V_\pm  & = \left[D_\pm, \chi \right] + [g B_\pm, \chi] =  \left[D_\pm + g B_\pm, \chi \right],
\end{align}
{\it i.e.} the field $V_\mu$ has consistent gauge transformation
properties corresponding to a gauge field. In the following we will
use the above parametrization of the gluonic field to expand the high
energy effective action for the quasi-elastic case around the
reggeized gluon field $A_+$ which we treat as a strong classical background
field $g A_+ \sim 1$. 

\subsection{The effective Lagrangian quadratic in $v_\mu$}
\label{sec:effect-lagr-quadr}

In the following we limit ourselves to the quasi-elastic case where
the Lagrangian contains only the induced terms corresponding to the
functional $W_-[v]$. The second set of induced  terms is  left aside for the
moment. This is sufficient to describe the interaction of a dilute
projectile with a target characterized by high parton densities in the
high energy limit, where the $A_+$ will couple through the reggeized
gluon propagator to color charges in the target.  To construct the
effective action for quasi-elastic processes, we use the following
parametrization of the gluonic field
\begin{align}
  \label{eq:para1}
  V^\mu(x) & =
 v^\mu(x)  +  \frac{1}{2} (n_-)^\mu  B_+[v_-]
\end{align}
and consider the following effective action for the quasi-elastic case
\begin{align}
  \label{eq:2}
  S_{\text{eff}}^{\text{q.e.}} & = S_{\text{QCD}} + S_{\text{ind.}}^{\text{q.e.}}
\end{align}
with
\begin{align}
  \label{eq:3}
   S_{\text{QCD}} & = \int d^4 x \left[  \tr  
 \left( \frac{1}{2} G_{\mu\nu}G^{\mu\nu} \right) 
+ \bar{\psi}(x) \left( i \slashed{D}  \right) \psi(x)  
\right]\,,
\end{align}
where $ G_{\mu\nu} = \frac{1}{g} \left[D_\mu, D_\nu \right]$ and
\begin{align}
  \label{eq:4}
  S_{\text{ind.}}^{\text{q.e.}} & = \int d^4 x\, \tr  \left( \left\{ T_-[v] - A_- (x) \right\}  \partial^2 A_+(x)\right]\,.
\end{align}
Keeping fields $A_+$ to all orders and expanding in quantum fluctuations $v_\mu$ and $\psi$, $\bar{\psi}$  to quadratic order we obtain
\begin{align}
  \label{eq:5}
   S_{\text{eff}}^{\text{q.e.}} & = \int d^4x \left[\mathcal{L}_0 + \mathcal{L}_1 - \tr\left( A_-\partial^2 A_+  \right)\right] + \mathcal{O}(v_\mu^3),
\end{align}
with the kinetic term of the gluonic  and quark field
\begin{align}
  \label{eq:l0}
  \mathcal{L}_0 & =  \tr \left( -v^\mu [g_{\mu\nu} \partial^2 - \partial_\mu \partial_\nu] v^\nu\right) + \bar{\psi} i \slashed{\partial} \psi
\end{align}
and the quadratic terms which describe interaction with the reggeized gluon field,
\begin{align}
  \label{eq:Lagrangian1X}
 \mathcal{L}_1 & = g\cdot  \bigg\{ \frac{i}{2}\bar{\psi} \slashed{n}_- A_+ \psi +  \tr \bigg[
 \partial_- v_\mu [ A_+, v^\mu] 
+  2 \partial_\mu v_- [v^\mu, A_+]   +
\notag \\
& \hspace{5cm}
  +    
  \partial^2 v_-   \left[\left(\frac{1}{\partial_-}v_- \right),  A_+ \right] -  v_- \left(\frac{1}{\partial_-}v_- \right) \partial^2 A_+ \bigg] \bigg\}\,.
\end{align}
Since we assume  that the reggeized gluon field couples to  high
partonic densities in the target, we have $g A_+ \sim 1 $; the term
$\mathcal{L}_1$ is therefore of the same order as $\mathcal{L}_0$. The
term  $\tr(A_- \partial^2 A_+ )$ provides the kinetic term of the
reggeized gluon field which is only needed to connect the $A_+$ field
to {\it e.g.} the target. 

\subsection{Parton-parton-reggeized gluon vertices}
\label{sec:part-part-regg}

The above Lagrangian $\mathcal{L}_1$ allows now for the straightforward determination of the quark-quark-reggeized gluon (QQR) and
gluon-gluon-reggeized gluon (GGR) vertex. Keeping an explicit dependence on
the reggeized gluon field, we find for quarks,
\begin{align}
  \label{eq:QQR}
  \parbox{2.8cm}{\includegraphics[width=2.8cm]{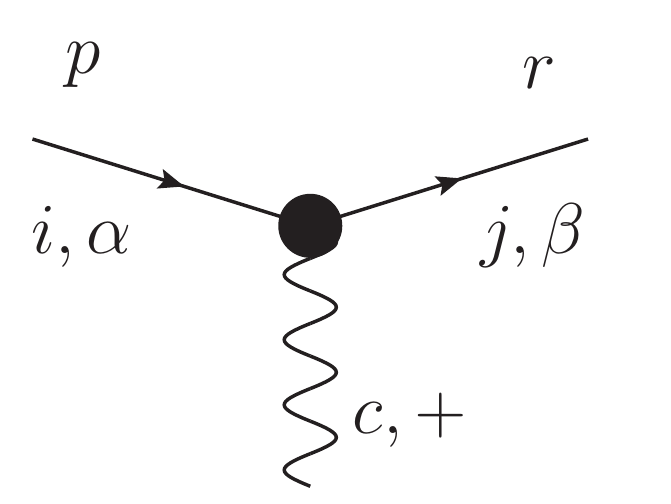}} & = -ig t^c_{ji} \Gamma_{\beta\alpha}(r,p) \int d^4z \, e^{-iz\cdot(p-r)} A^c_+(z), &   \Gamma_{\beta\alpha}(r,p) & = -\frac{1}{2} \slashed{n}^+_{\alpha\beta}\,,
\end{align}
which coincides with the expression used {\it e.g.} in
\cite{quarkjet}. For gluons one obtains instead
\begin{align}
  \label{eq:GG}
  \parbox{2.8cm}{\includegraphics[width=2.8cm]{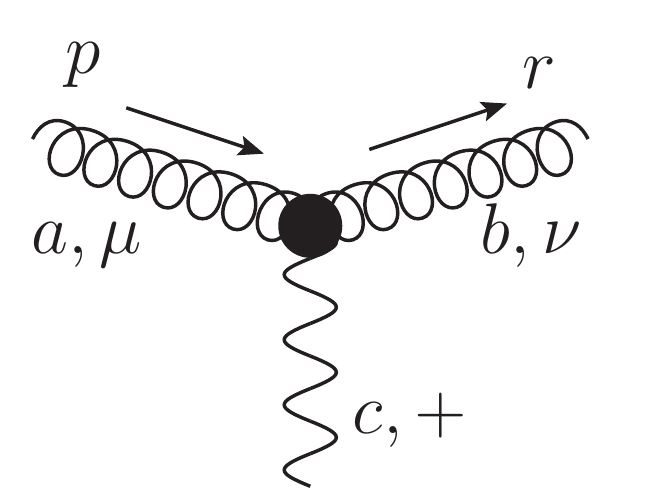}} & = -ig T^c_{ba} \Gamma^{\nu\mu}(r,p)  \int d^4z \, e^{-iz\cdot(p-r)} A^c_+(z),  \notag \\
   \Gamma_+^{\nu\mu}(r, p) & = p^+ g^{\mu\nu} - (n^+)^\mu p^\nu - (n^+)^\nu r^\mu  + \frac{r \cdot p}{p^+}  (n^+)^\mu (n^+)^\nu \notag \\
 & = p^+ g^{\mu\nu}_\perp  - (n^+)^\mu {\bm p}^\nu - (n^+)^\nu {\bm r}^\mu 
 - \frac{{\bm r} \cdot {\bm p}}{p^+}  (n^+)^\mu (n^+)^\nu\,,
\end{align}
with $T^c_{ab} = -if^{abc}$. Since $\partial_-A_+ = 0$, the integral
over $z$ yields for both vertices a $\delta(p^+ - r^+)$. We note that the above
GGR-vertex was already obtained in \cite{Lipatov:1995pn}; it differs
from the GGR-vertex obtained in {\it e.g.} 
\cite{gluonjet,Antonov:2004hh}, which is derived using  the direct transition
vertex Fig.~\ref{fig:3}.a. The above GGR vertex obeys the
following important properties: at first one finds current
conservation on the level of the vertex, even if the  the second gluon  is not real   and/or does not 
carry physical polarization, 
\begin{align}
  \label{eq:9}
 r_\nu \cdot \Gamma^{\nu\mu}_+(r,p) & = 0 =    \Gamma^{\nu\mu}_+(r, p) \cdot p_\mu\,.
\end{align}
A disadvantage of the above vertex, already noticed in
\cite{Lipatov:1995pn} is that the term $p\cdot r/p^+$ is in potential
conflict with the Steinmann-relations \cite{Steinmann}, since it may
yield individual Feynman diagrams which contain singularities in
overlapping  channels {\it e.g.} the $s$ and the $t$-channel. Nevertheless,
since this vertex is obtained from a shift in the gluonic field from
an effective action which explicitly obeys the Steinmann-relations,
the terms which potentially violate the Steinmann relations should
cancel for physical quantities. Application of this vertex to the
calculation of physical observables should be therefore safe.  Apart
from the above relation, this GGR-vertex also obeys
\begin{align}
  \label{eq:7}
n^+_\nu \cdot \Gamma^{\nu\mu}_+(r,p) & = 0 =    \Gamma^{\nu\mu}_+(r, p) \cdot n^+_\mu \,,
\end{align}
as well as
\begin{align}
  \label{eq:8}
 \Gamma^{\nu\alpha}_+(r,k)\cdot  (-g_{\alpha\alpha'} )\cdot  \Gamma^{\alpha'\mu}_+(k,p) & =   -p^+   \Gamma^{\nu\mu}_+(r,p)\,.
\end{align}
Identical properties hold for the QQR-vertex,
\begin{align}
  \label{eq:10}
\Gamma_{\beta\gamma'}(r,p) \slashed{n}_{\gamma'\gamma} &= 0   = 
  \slashed{n}_{\beta\beta'}  \Gamma_{\beta'\gamma}(r,p) \,,
   \notag \\
\Gamma_{\beta\gamma}(r,p) \slashed{k}_{\gamma\gamma'} \Gamma_{\gamma'\alpha}(r,p) & = -p^+ \Gamma_{\beta\alpha}(r,p)\,.
\end{align}

\subsection{Properties of the reggeized gluon field}
\label{sec:prop-regg-gluon}
The last two properties Eq.~\eqref{eq:8} and Eq.~\eqref{eq:10} are of
high importance to arrive at  a  summation of the reggeized gluon
field to all orders. Before  addressing this task,  we first  recall the following property of the reggeized gluon field, 
\begin{align}
  \label{eq:7}
 \partial_- A_+ (x) & = 0, &   A_+ (x)& = A_+ (x_0^-, {\bm x}, x^+) \notag \\
\partial_+ A_- (x) & = 0, &   A_- (x)& = A_- (x_0^+, {\bm x}, x^+) \,,
\end{align}
with a $x_0^\pm$ a constant which is common to all $A_+$ fields; since
the scattering amplitude dependes by Lorentz invariance not on
absolute space-time values, this constant can be conveniently set to
$x_0^\pm = 0$. To keep the presentation as general as possible, we
keep in the following however the dependence on $x_0^\pm$ and set it
only to zero when comparing to other approaches. We further recall
that the propagator of the reggeized gluon field, Fig.~\ref{fig:3}.b,
which connects clusters significantly separated in rapidity, comes
with a purely transverse denominator. The corresponding configuration
space propagator is therefore in four dimensions given by
\begin{align}
  \label{eq:11}
  \langle A_+(x)A_-(y) \rangle & = \int \frac{d^4 q}{(2 \pi)^4} e^{-iq\cdot(x - y)}\frac{2i}{{\bm q}^2} \notag \\
& = \frac{1}{2} \int  \frac{d^2 {\bm q}}{(2 \pi)^2}
\int \frac{d q^+}{2 \pi} e^{-iq^+(x_0^- -  y^-)/2}
\int \frac{d q^-}{2 \pi} e^{-iq^- (x^+ - x_0^+) /2}
 e^{i{\bm q}\cdot({\bm x} - {\bm y})}\frac{2i}{{\bm q}^2} 
\notag \\
& =  4\delta( y^- - x_0^-)\delta(x^+ - x^+_0)  \cdot \int \frac{d^2 {\bm q}}{(2 \pi)^2}
e^{i{\bm q}\cdot({\bm x} - {\bm y})}\frac{i}{{\bm q}^2}  .
\end{align}
The four dimensional reggeized gluon propagator can therefore be
interpreted as  the propagator of a two-dimensional
reggeized gluon field $\alpha({\bm z})$, together with corresponding delta functions, 
\begin{align}
  \label{eq:13}
  \langle A_+(x)A_-(y)\rangle & = 4\delta(x^+-x^+_0) \delta(y^- - x^-_0)  \cdot \langle \alpha({\bm x}) \alpha (\bm y)\rangle,
\end{align}
with
\begin{align}
  \label{eq:21}
\langle \alpha({\bm x}) \alpha (\bm 0)\rangle
&= \int \frac{d^2 {\bm q}}{(2 \pi)^2}
\frac{i e^{i{\bm q}\cdot({\bm x})}}{{\bm q}^2} 
.
\end{align}
The result then suggests to parametrize the reggeized gluon field as :
\begin{align}
  \label{eq:10}
   A_+ (x)& = 2 \cdot \alpha ({\bm x}) \delta(x^+ - x^+_0)\,,
\end{align}
where the factor of two appears due to the chosen convention for
light-cone directions.  We note that such a parametrization is
commonly used in calculations within the CGC-formalism, see {\it e.g.}
\cite{Balitsky:1995ub, jimwlk1,jimwlk2,jimwlk6,jimwlk8}, with
$x_0^+ = 0$.  This treatment of the reggeized gluon field is possible,
since the fields $A_\pm$ are within the effective action to be treated
as external classical fields for individual rapidity clusters, while
they only connect to other clusters through the above reggeized gluon
propagator.

\subsection{All order summation of the reggeized gluon fields}
\label{sec:all-order-summation}

To sum up the interaction of partons with reggeized gluon fields to
all orders in $\alpha_s$, it is necessary to  determine the free gluon
propagator of the quantum fluctuations $v^\mu$, which requires fixing a gauge  following the usual Faddeev-Popov procedure.  While the following
discussion will be based on  covariant gauge,  we will also comment on the corresponding results obtained in axial light cone gauge  with the free propagators  given by the usual expressions
\begin{align}
  \label{eq:free}
  \tilde{G}_{\text{cov.},\mu\nu}^{(0),ab}(k) & = \delta^{ab}\tilde{D}_0(k)  \left[- g_{\mu\nu} + (1 - \xi) \frac{k_\mu k_\nu}{k^2}\right] = \delta^{ab}d_{\mu\nu}(k, \xi) \tilde{D}_0(k)\,, \notag \\
 \tilde{G}^{(0),ab}_{\text{l.c.}, \mu\nu}(k) & =\delta^{ab} \tilde{D}_0(k)  \left[- g_{\mu\nu} +\frac{k_\mu (n^+)_\nu +  (n^+)_\mu  k_\nu}{k \cdot n^+}\right] = \delta^{ab} d_{\text{l.c.},\mu\nu}(k, n^+) \tilde{D}^{(0)}(k)\,,
\end{align}
with
\begin{align}
  \label{eq:G0}
\tilde{D}^{(0)}(k) & =    \frac{i}{k^2 + i\epsilon}.
\end{align}
If not denoted otherwise, we will in the following always  use covariant gauge. 
For the quark propagator one finds the usual expression
\begin{align}
  \label{eq:12}
  \tilde{S}^{(0)}_{F}(k) & = \slashed{k}   \tilde{D}^{(0)}(k)\,.
\end{align}
Due to the properties  Eq.~\eqref{eq:9}, Eq.~\eqref{eq:7} , connecting two  GGR vertices with a gluon propagator, the polarization tensor of the latter reduces always to $-g_{\mu\nu}$, since all other terms are set to zero. Using further the properties Eqs.~\eqref{eq:8} and \eqref{eq:10}, the interaction of $n$ reggeized gluons with a quark or gluon reduces to  essentially to
\begin{align}
  \label{eq:15}
  &  \prod_{i=1}^n \int d z_i^4   \prod_{j=1}^n \int  \frac{d^4 k_j}{(2 \pi)^4} \,  
 (-k_1^+) D_0(k_1)  e^{i k_1\cdot (z_1 - z_{2})}\ldots (-k_{n-1}^+) D_0(k_{n-1})   e^{i k_{n-1}\cdot (z_{n-1} - z_{n})}    \notag \\
&  \hspace{5cm}  e^{-ip\cdot z_1} \left(-igA_+(z_n)\right) \ldots \left(-ig A_+(z_1)\right)  e^{ ir\cdot z_n}  \notag \\
& = 
-2 \pi \delta( p^+ - r^+) e^{-i x_0^+(p^- - r^-)} \int d^2 {\bm z} e^{i {\bm z} \cdot ({\bm p} - {\bm r})}  \notag \\ 
&  \hspace{2.5cm} \bigg[ \theta(p^+)   \mathrm{P} \left( \frac{-g}{2} \right)^n  \int \prod_{i=1}^n dz^+_i \tilde{A}_+(z_i) 
 -\theta(-p^+)  \overline{\mathrm{P}} \left( \frac{g}{2} \right)^n  \int \prod_{i=1}^n dz^+_i \tilde{A}_+(z_i) 
\bigg]\,.
\end{align}
 To arrive at the above identity, we used the property
Eq.~\eqref{eq:10}.  $A_+ = - it_{ji}^c A_+^c$ are reggeized gluon
fields in the fundamental representation for quarks while gluons require
$A_+ \to \tilde{A}_+= - iT_{ba}^c A_+^c$ {\it i.e.} reggeized gluon fields in the
adjoint representation. (Anti-)path
ordering of color matrices is as usually defined as
\begin{align}
  \label{eq:pathordering}
  \mathrm{P} A_+(z_n^+, {\bm z} ) \cdots A_+(z^+_1,  {\bm z}) & \equiv   A_+(z^+_n, {\bm z}) \cdots A_+(z^+_1, {\bm z}) \theta(z^+_n - z_{n-1}^+)  \ldots \theta(z_2^+  - \ldots z^+_1) \, \notag \\
 \overline{ \mathrm{P}} A_+(z_n^+, {\bm z} ) \cdots A_+(z^+_1,  {\bm z}) & \equiv   A_+(z^+_1, {\bm z}) \cdots A_+(z^+_n, {\bm z}) \theta(z^+_n - z_{n-1}^+)  \ldots \theta(z_2^+  - \ldots z^+_1)  .
\end{align}
Summing finally over the number of reggeized gluons, one obtains for gluons the following effective  vertex which sums up the interaction with an  arbitrary number of reggeized gluon fields,
\begin{align}
  \label{eq:finally}
\parbox{4cm}{\includegraphics[width=4cm]{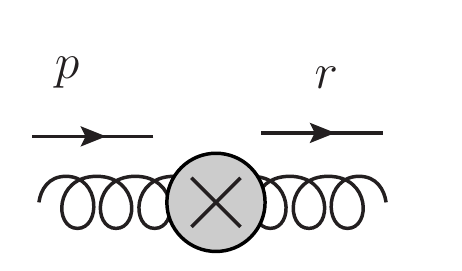}} & = 
\tau_{G,\nu\mu}^{ab}(p, -r) =  
  - 4 \pi \delta(p^+ - r^+)  \Gamma_{\nu\mu}(r,p)  e^{-i x_0^+(p^- - r^-)}
\notag \\ & \hspace{-1cm}
 \cdot \int d^2 {\bm z} e^{i {\bm z} \cdot ({\bm p} - {\bm r})}   
\bigg[ \theta(p^+)  \left[ U^{ba}({\bm z}) - \delta^{ab} \right]-  \theta(-p^+) \left[  [U^{ba}({\bm z})]^\dagger  - \delta^{ab} \right]\bigg].
\end{align}
For quarks one finds,
\begin{align}
  \label{eq:quark_vertex}
 \parbox{4cm}{\includegraphics[width=4cm]{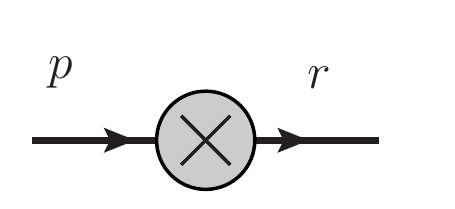}} & = \tau_{F}(q,-r)=
  2 \pi \delta(p^+ - r^+) \slashed{n}^+  e^{-i x_0^+(p^- - r^-)}  
\notag \\
& \cdot  \int d^2 {\bm z} e^{i {\bm z} \cdot ({\bm p} - {\bm r})}
\bigg[ \theta(p^+)  \left[ W({\bm z}) - 1 \right]-  
\theta(-p^+) \left[  [W({\bm z})]^\dagger  - 1 \right]\bigg]\,.
\end{align}
To write down the above expressions,  we introduced Wilson lines in the adjoint
\begin{align}
  \label{eq:Uab}
   U^{ab}({\bm z}) & =    \mathrm{P} \exp \left(-\frac{g}{2}\int_{-\infty}^\infty dz^+ \tilde{A}_+ \right), &   \tilde{A}_+ & = -iT^c_{ab} A^c_+\,,
\end{align}
and the fundamental representation
\begin{align}
  \label{eq:W}
  W({\bm z}) & =   \mathrm{P} \exp \left(-\frac{g}{2}\int_{-\infty}^\infty dz^+ {A}_+ \right), &   {A}_+ & = -it^c_{ij} A^c_+\,.
\end{align}
In contrast to the notation used in  \cite{Hentschinski:2011xg,Ayala:2017rmh} and elsewhere,  we use here the letter $W$ to denote the  Wilson line in the fundamental representation to avoid confusion with the gluonic field in the effective action. The above expressions Eq.~\eqref{eq:finally} and Eq.~\eqref{eq:quark_vertex} are one of the central results of this paper. 

\section{Comparison with  expressions in the literature}
\label{sec:literature}

At this stage it is necessary to compare the result derived from
Lipatov's high energy effective action with the conventional quark and
gluon propagators in the presence of a background field used in the
literature.

\subsection{Comparison with propagators in the presence of a background field}
\label{sec:comp-with-prop}

 Corresponding resummed propagators are within the effective action
 now easily obtained.  Using Eqs.~\eqref{eq:finally} and
 \eqref{eq:quark_vertex} one finds for the resummed quark ($S_F$) and
 gluon ($G$) propagators:
\begin{align}
\label{eq:prop-action}
S_{F} (p,q) &=
 S_{F}^{(0)} (p) (2 \pi)^4 \delta^{(4)}(p - q) \,+\,
S_{F}^{(0)}(p)  \, \cdot \,  \tau_{F} (p,q) \, \cdot  \, S_{F}^{(0)} (q)\, ,
\notag \\
G_{\mu\nu}^{ad} (p,q) &= G^{(0),ab}_{\;\;\mu\nu} (p) (2 \pi)^4 \delta^{(4)}(p - q) \, + \,  G^{(0),ab}_{\;\;\mu\alpha} (p) \, \cdot \,  \tau_{G}^{\alpha\beta,bc} (p,q)  \,  \cdot \,   G^{(0), cd}_{\;\;\beta \nu} (q) \, ,
\end{align}
where for the moment we do not specify the gauge  of the free gluon
propagators. These expression are now to be compared with propagators
obtained from treating the target as a background field in light-cone
gauge $b \cdot n_- = 0$ with the only non-zero component
\begin{align}
\label{eq:bplus}
  b_+(x^+, {\bm z}) = \delta(x^+)\beta({\bm z}),
\end{align}
while $b^\mu_\perp = 0$. Using the Fourier transform of corresponding
counter parts in configuration space, see {\it e.g.}
\cite{McLerran:1994vd} one finds in momentum space (see {\it e.g.}
\cite{Ayala:2017rmh} for expressions used in 
a recent calculation),
\begin{align}
\label{eq:prop-bg}
S^{[b]}_{F} (p,q) &=
 S_{F}^{(0)} (p) (2 \pi)^4 \delta^{(4)}(p - q) +
S_{F}^{(0)}(p)  \, \cdot \,  \tilde{\tau}_{F} (p,q) \, \cdot  \, S_{F}^{(0)} (q)\, ,
\notag \\
G_{\mu\nu}^{[b],ad} (p,q) &= G^{(0),ab}_{\text{l.c.},\mu\nu} (p) (2 \pi)^4 \delta^{(4)}(p - q) + G^{(0),ab}_{\;\;\mu\alpha} (p) \, \cdot \,  \tilde{\tau}_{G}^{\alpha\beta,bc} (p,q)  \,  \cdot \,   G^{(0), cd}_{\text{l.c.},\beta \nu} (q) \, ,
\end{align}
where the gluon propagator is now restricted to $v \cdot n_- = 0$
light-cone gauge. The superscript `$[b]$' indicates that these
propagators have been derived using the background field in light-cone
gauge and not the reggeized field $A_+$. One has
\begin{align}
  \label{eq:quarkinteractionCGC}
\tilde{\tau}_{F}(p,-q)   & = 2 \pi  
 \delta(p^+ - q^+) ~  \slashed{n}^+
\notag \\
&  \times \int d^{2} {\bm z} e^{i{\bm  z} \cdot ({ \bm p} - { \bm q})}
 \left\{\theta(p^+) \big[ W[b]({\bm z}) -1 \big]
-
\theta(-p^+) \big[ W[b]^\dagger({\bm z}) -1 \big]
  \right\} \\
\label{eq:gluoninteractionCGC}
 \tilde{\tau}_{G,\nu\mu}^{ab}(p,q)&=
2 \pi \delta(p^+ - q^+) ~ ( - 2 p^+ g_{\nu\mu})
\notag \\
&  \times
\int d^{2} {\bm z} e^{i{\bm z} \cdot ({\bm p} - {\bm q})}
 \left\{\theta(p^+) \big[ U^{ab}[b]({\bm z}) -1 \big]
-
\theta(-p^+) \big[ \left(U^{ab}[b]\right)^\dagger({\bm z}) -1 \big]
  \right\},
\end{align}
with Wilson lines in fundamental ($W$) and adjoint ($U$)
representation
\begin{align}
  \label{eq:wilson}
  W[b](\bm z) &  = \mathrm{P} \exp \left( -\frac{g}{2} \int\limits_{-\infty}^\infty
 d x^+ b^{-,c}(x^+, {\bm z})t^c  \right),
&
b^{-}(x^+, {\bm z}) & = -i b^{-,c}(x^+, {\bm z})t^c
\notag \\
 U[b](\bm z) & = \mathrm{P} \exp \left(- \frac{g}{2} \int\limits_{-\infty}^\infty
d x^+ b^{-,c}(x^+, {\bm z})T^c \right),
&
\tilde{b}^{-}(x^+, {\bm z}) & = -i b^{-,c}(x^+, {\bm z})T^c \,.
\end{align}
Leaving aside potential differences in the Wilson lines, to which we
will turn in Sec.~\ref{sec:wilsonlines}, one observes that both quark
propagators agree directly with each other (if one sets $x_0^+ = 0$).
To carry out a similar comparison for the gluon, we consider first the
case where the external free propagators in Eq.~\eqref{eq:prop-action}
are taken in $v \cdot n_- = 0$ light-cone gauge. Since
$d_{\text{l.c.}}^{\mu\nu}(p, n^+) n^+_\nu = 0 =
d_{\text{l.c.}}^{\mu\nu}(r, n^+) n^+_\mu$,
all terms in the vertex $\Gamma^{\nu\mu}(r,p)$ which contain a
$n^+_\mu$ or $n^+_\nu$ vanish. One therefore remains with the
$2 p^+ g_{\mu\nu}$ term only which is precisely the term used in
Eq.~\eqref{eq:gluoninteractionCGC}. Both expression therefore agree for  $x_0^+ = 0$.
We further note that both the light-cone gauge polarization tensor and
the GGR-vertex can be factorized into the products of a `left' and
`right' tensor,
\begin{align}
  \label{eq:lc_connect1}
 c_L^{\mu\alpha}(p, n^+) & =   \left( g^{\mu\alpha} - \frac{(n^+)^\mu p^\alpha}{p \cdot n^+}  \right) 
&
 c_R^{\alpha\nu}(r, n^+)   & =  \left( g^{\alpha\nu} - \frac{ r^\alpha (n^+)^\nu}{r \cdot n^+}  \right)\,, 
\end{align}
where
\begin{align}
  \label{eq:moreconnect}
   \Gamma^{\mu\nu} & = p^+ c_L^{\mu\alpha} (p, n^+)  c_R^{\alpha\nu}(r, n^+),
\end{align}
and
\begin{align}
  \label{eq:16}
  d^{\mu\nu}(p, n^+) & = c_R^{\mu\alpha}(p, n^+) (-g_{\alpha\beta}) c_L^{\beta\nu}(p, n^+).
\end{align}
This property allows to establish on a diagrammatic level how the
vertex $\Gamma^{\mu\nu}$ can build up from properly factorizing the
numerator of the light-cone gauge gluon propagator and absorbing them into the vertex; the information
contained in Eq.~\eqref{eq:prop-action} and \eqref{eq:prop-bg} is
therefore in this sense identical. It is an interesting note aside that
a similar mechanism has been used in the construction of a certain
projector in \cite{Gituliar:2015agu}.

\subsection{Comparison of Wilson lines and the definition of the reggeized gluon}
\label{sec:wilsonlines}

In the following we attempt a somewhat detailed comparison between the
Wilson lines in the reggeized gluon field $A_+$, arising from
Lipatov's high energy effective action, and Wilson lines in the
background field $b_+$, frequently encountered in CGC calculation in
light-cone gauge. While we find that the interpretation of these
Wilson lines differs, we would like to stress that for the calculation
of correlators in the dilute quasi-elastic region, {\it i.e.}
perturbative forward scattering in the presence of a strong background
field (reggeized gluon or light-cone gauge), both formalism are
equivalent; the only difference is that the effective action allows
use of arbitrary gauges\footnote{Nevertheless we would like to stress
  that calculation based on the background field in light-cone gauge
  allow at least in principle also for the use of different gauges for
  the gluon fluctuations.}. The difference lies therefore mainly in
the interpretation of the background field,  {\it i.e.}  the coupling
to color sources in a different rapidity cluster. At first both Wilson lines appear to resum identical
fields; Eq.~\eqref{eq:10} and Eq.~\eqref{eq:bplus} take identical
forms. Obviously one has for a Wilson line of a generic gluonic field
$V_+$,
\begin{align}
  \label{eq:Wperm}
    W[V](x)    &=  \mathrm{P} \exp\left(-\frac{g}{2} \int\limits_{-\infty}^\infty  dx^+ V_+(x) \right)
=
 \sum_{n=0}^\infty  \frac{ \left({-g} \right)^n }{2^n n!} \int  \prod_{i=1}^n d x_i^+
\notag \\ & \hspace{2cm}
\bigg[  V_+(x_1)\dots V_+(x_n) \theta(x_1^+ - x_2^+) \ldots \theta(x^+_{n-1} - x_n^+) 
+ \text{permutations} \bigg].
\end{align}
If now
$V_+(x) = A_+(x) = -2i\delta(x^+ - x^+_0) \alpha^a({\bm x})t^a$,  the
permutations of the fields $A(x_i)$, $i = 1, \ldots, n$ are all identical (since their $x^+$ dependence is identical) and we arrive directly at
\begin{align}
  \label{eq:Wperm}
    W[A](x)  &=   \sum_{n=0}^\infty  \frac{1}{n!} \left(\frac{-g}{2} \right)^n   \prod_{i=1}^n \int d x_i^+  A_+(x_1)\dots A_+(x_n)
 \notag \\ &   \hspace{3cm}
   \left[\theta(x_1^+ - x_2^+) \ldots \theta(x^+_{n-1} - x_n^+) + \text{permutations} \right] \notag \\
&
 = \sum_{n=0}^\infty  \frac{1}{n!} \left(\frac{-g}{2} \right)^n   \prod_{i=1}^n \int d x_i^+ 
  A_+(x_1)\dots A_+(x_n) 
 = e^{{ig}\alpha^a({\bm x}) t^a },
\end{align}
We therefore obtain a simple matrix exponential. Formally, also the
choice $V_+(x) = b_+(x) = -i\delta(x^+) \beta^a({\bm x}, x^-)t^a$
leads obviously to the same result. In the literature such an
interpretation is however usually avoided, by treating the contracting
of the $x^+$-dependence to delta-like support as an approximation
which applies to the calculation of correlators in the background
field, while the $b_+$ itself is ordered in the $x^+$ coordinates. see
{\it e.g.} \cite{jimwlk8}.
\\

While the precise interpretation used is irrelevant for the
calculation of correlators in the presence of a background field, the
difference becomes striking once correlators of the background field
with {\it e.g.} color charges in a rapidity cluster significantly
separated in rapidity are considered (``the dense target'').  Vertices
which describe the interaction of the Wilson line with $n$-reggeized
gluons fields come with purely symmetric color tensors, since the
precise ordering of fields is irrelevant. For the gluonic field
$b_+(x)$ such a result is not acceptable, since one would miss the
corresponding anti-symmetric and mixed symmetry correlators.  Within
the effective action, the interaction with these color charges does
not occur directly through the reggeized gluon field, but through the
induced vertices Fig.~\ref{fig:3} and corresponding higher order
vertices. Following the treatment in \cite{Hentschinski:2011xg},
theses vertices carry only anti-symmetric color tensors (corresponding
to a combination of anti-commutators of SU$(N_c)$
generators). Combining these induced vertices with the symmetric $m$
reggeized gluon state to construct a `Wilson-line-$n$ gluon' vertex
($n \geq m$), where the coupling to the Wilson line is always mediated
by at least one reggeized gluon, one recovers the complete symmetry
structure. For a pedagogic presentation for the case up to three
gluons we refer to Appendix \ref{sec:multi-gluon-exchange}; see also
the discussion in \cite{Hentschinski:2009zz}.
\\

At this point we would like to return to a proposal made in \cite{Caron-Huot:2013fea} for the definition of the reggeized gluon from Wilson-lines in the Balitsky-JIMWLK formalism. There it has been proposed to define the reggeized gluon $R^a({\bm z})$ as the logarithm of the adjoint Wilson line,
\begin{align}
  \label{eq:17}
  R^a({\bm z}) &\equiv  \frac{1}{g N_c} f^{abc} \log U^{bc}({\bm z})\,.
\end{align}
Using the  above results, one finds directly for the results obtained from Lipatov's high energy effective action,
\begin{align}
  \label{eq:18}
   R^a({\bm z}) & =   \frac{1}{g N_c} f^{abc} \left[{ig} \alpha^d({\bm z})T^d_{bc} \right] = \alpha^a({\bm z}) = \frac{1}{2} \int dx^+ A_+^a(x^+, {\bm z}),
\end{align}
{\it i.e.} the definition of the reggeized gluon of
\cite{Caron-Huot:2013fea} coincides with the reggeized gluon field of
Lipatov's effective action, once this field is integrated over the
corresponding light-cone coordinate\footnote{At least within the high energy effective action,  a definition based on the  Wilson lines in the fundamental representation would be equally possible, {\it i.e.} $R^a({\bm z}) = \frac{2}{ig}\tr(t^a \log [V({\bm z})]) = \alpha^a({\bm z})$}.

\section{Balitsky-JIMWLK evolution}
\label{sec:balitsky-jimwlk-evol}

In the following we demonstrate that the high energy evolution of
Wilson lines of reggeized gluons (obtained within the high energy
effective action) leads directly to the leading order Balitsky-JIMWLK
evolution equation. Even though this is expected, given the
coincidence in the resummed gluon and quark propagators, this provides
an important consistency check, in particular for future calculation
of CGC-observables.  We will then investigate the question whether
integrating out quantum fluctuations of a general ensemble of Wilson
lines gives indeed rise to the Balitsky-JIMWLK evolution equation.
\\

Within Lipatov's high energy effective action, the determination of
high energy evolution requires in general the high energy effective
action for `central-rapidity' processes, {\it i.e.} the effective
action which contains both $A_-$ and the $A_+$ reggeized gluon fields
and corresponding induced vertices. For the discussion of dense-dilute
collision the decomposition provided by the effective action for
central rapidities is however not very efficient; the additional set
of induced vertices provides a certain color decomposition of
amplitudes which describe gluon production from a multi-reggeized
gluon exchange. While it has been demonstrated at the level of the
scattering amplitude for four-reggeized gluon exchange that after a
certain reshuffling of terms the $2-4$ reggeized gluon vertex (triple
Pomeron vertex) arises from the high energy effective
action\cite{Hentschinski:2009zz} (which at the same time can be shown
to arise as well from Balitsky-JIMWLK evolution
\cite{Bartels:2004ef}), the calculation is rather cumbersome. While
the reformulation of the effective action provided in
Sec.~\ref{sec:parametrize} already provides a first simplification, it
is easier to recover the Balitsky-JIMWLK evolution equation from the
quantum fluctuations of the quasi-elastic Lagrangian.  For an ensemble
of Wilson lines the latter are directly proportional to the high
energy divergence, without the need to drop any finite terms. We hope
to return to the description which uses the
high energy effective action for central rapidity processes in a future publication. \\

For the following discussion it sufficient to consider Wilson lines in the fundamental representation. While adjoint Wilson lines can be rewritten in terms of fundamental Wilson lines using the well-known relation
\begin{align}
  \label{eq:4}
  U^{ab}({\bm z}) & = 2 \tr \left[ t^a W({\bm z})t^b  W^\dagger({\bm z})\right]\,,
\end{align}
the hermitian conjugate of a fundamental Wilson lines follows
trivially from the discussion of the fundamental Wilson line. We will
therefore consider the quantum fluctuations of an ensemble of $n$
fundamental Wilson lines in the reggeized gluon fields,
\begin{align}
  \label{eq:5}
  W[A_+]({\bm z}_1) \otimes \ldots \otimes  W[A_+]({\bm z}_n).
\end{align}

\subsection{Feynman rules for quantum fluctuations of a Wilson line}
\label{sec:fr}
\begin{figure}[th]
  \centering
\parbox{3.5cm}{\includegraphics[height=1.5cm]{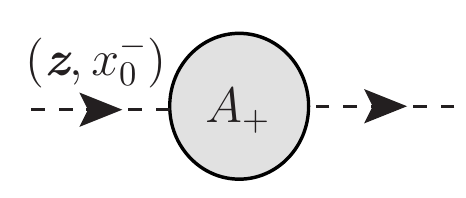}} $  =
                                 \displaystyle  W[A_+]({\bm z}, x_0^-) $,
$\qquad$
 \parbox{2cm}{\includegraphics[width=2cm]{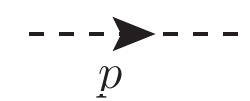}} 
                            $ =    \displaystyle  \frac{i}{p^- + i\epsilon} $,
 \\ 
\parbox{7cm}{\center (a)} \parbox{5cm}{\center (b)} \\
\parbox{3.5cm}{\includegraphics[width=3.5cm]{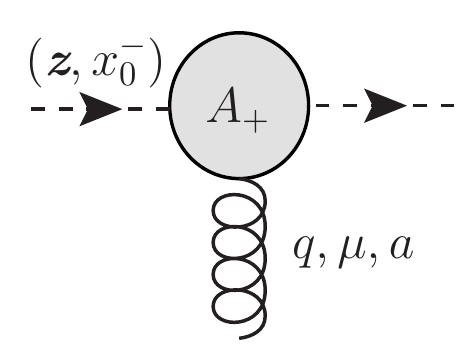}} $ = \displaystyle  \frac{g}{q^+}  (n^+)^\mu e^{-iq^+ x_0^-/2 + i {\bm q} \cdot {\bm z}} \cdot  \bigg[ W[A]({\bm z}, x_0^-), t^a \bigg] $, \\
\parbox{10cm}{\center (c)} \\
 \parbox{4cm}{\includegraphics[width=3.5cm]{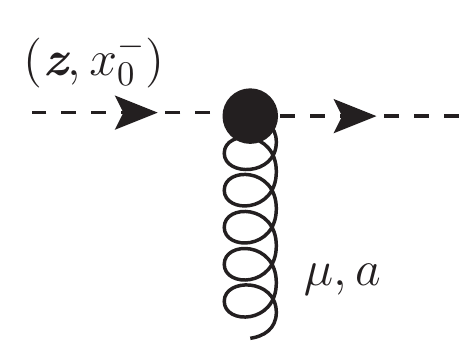}}    $   = \displaystyle  ig t^a (n^-)^\mu e^{-iq^+ x_0^-/2 + i {\bm q} \cdot {\bm z}} .$ \hspace{2.5cm} $\,$ \\ 
\parbox{10cm}{\center (d)}
  \caption{\small Feynman rules for the calculation of quadratic
    fluctuations of the Wilson lines for covariant or $v_-=0$ gauge. Note that the the Wilson-line-gluon vertex (d) conserves momentum as usually, while four momenta are not conserved at the vertices (a) and (c). Momenta which are not fixed by external momenta are understood to be integrated over with the measure $d^4 p/(2 \pi)^4$}. 
  \label{fig:feynman_wilson}
\end{figure}
Integrating out the quantum fluctuations $v^\mu$ is most easily
achieved, if one supplements the effective action with an auxiliary
complex 1-dimensional scalar field,
$\varphi = \varphi(x^+, {\bm z}, x_0^-)$ where ${\bm z}, x_0^-=0$ are
constant for the dynamics of the scalar field. The field is charged in
the fundamental representation of $SU(N_c)$ and transforms under gauge
transformations as
\begin{align}
  \label{eq:gauge_scalar}
  \delta_L \varphi & = - \chi_L \varphi.
\end{align}
The 1-dimensional gauge invariant action  of this field, which describes interaction with the gluonic field, is given by
\begin{align}
  \label{eq:Lagrangian_scalar}
  S[\varphi, V] & =
   \int d x^+  \varphi^\dagger \left[ i \partial_+  + i g v_+ \right]\varphi \,,
\end{align}
where all fields are taken at fixed $({\bm x}, x_0^-)$. One obtains in
a straightforward manner for the propagator of this scalar field
\begin{align}
  \label{eq:propagator}
 \left \langle x^-\left |  \frac{1}{1 + \frac{g}{\partial_+ +  \epsilon} V_+} \frac{1}{\partial_+ + \epsilon} \right |y^- \right \rangle & = \mathrm{P} \exp \left(\frac{-g}{2} \int_{y^+}^{x^+} d z^+  v_+ \right).
\end{align}
 As a next step  we use the parametrization Eq.~\eqref{eq:para1} of
 the gluonic field and  limit ourselves to terms
quadratic in the quantum fluctuation.  Limiting ourselves further  to
covariant or $v_-=0$ gauges, the following simplified shift is sufficient\footnote{Covariant gauge requires correlators of $v_-$ and $v_+$ fields as well as two $v_+$ fields; the correlator of two $v_-$ vanishes on the other hand. $v_-=0$ gauge requires on the other hand only the correlator of  two $v_+$ fields}, 
\begin{align}
  \label{eq:shift_new}
  v^\mu & \to  V^\mu = v^\mu + 
\frac{1}{2}(n_-)^\mu  \left( A_+ + [A_+, \frac{g}{\partial_-}v_-] \right) + \mathcal{O}(v_-^2).
\end{align}
Expanding our expressions around the background field $gA_+\sim 1$,
the shifted action is given by
\begin{align}
  \label{eq:Lagrangian_scalar_shift}
  S[\varphi, A_+,  v] & = \int d x^+  \varphi^\dagger  \left[ i \partial_+  + i g\left(  v_+ + A_+ + [A_+, \frac{g}{\partial_-}v_-]  \right)\right] \varphi\, .
\end{align}
The resulting set of Feynman rules necessary for the calculation of $\mathcal{O}(g^2)$ corrections within covariant and/or $v_-=0$ gauge are then summarized in Fig.~\ref{fig:feynman_wilson}.

\subsection{Calculating quantum fluctuations}
\label{sec:int2}

Since we require only fluctuations up to quadratic order, it is
sufficient to consider the   correlator of two Wilson lines at 1-loop.
The non-zero diagrams for self-energy type corrections to one Wilson
line are given by
\begin{align}
  \label{eq:30}
  \parbox{4cm}{\includegraphics[height=2cm]{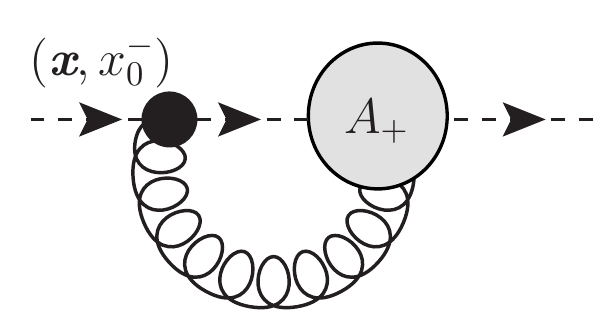}} + 
  \parbox{4cm}{\includegraphics[height=2cm]{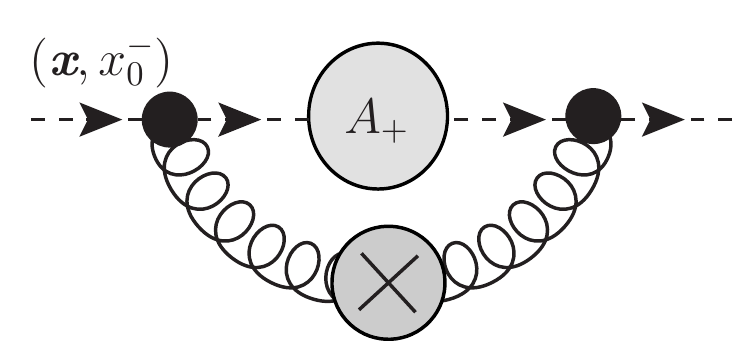}} + 
  \parbox{4cm}{\includegraphics[height=2cm]{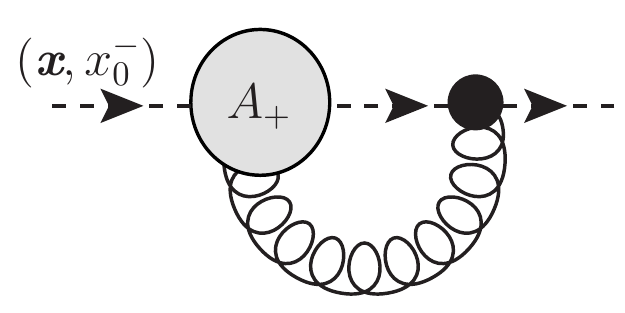}.} 
\end{align}
For interactions between 2 Wilson lines, evaluation of the following
diagrams is sufficient (the remaining diagrams can be deduced from symmetry),
\begin{align}
  \label{eq:diag_2lines}
  \parbox{4cm}{\includegraphics[height=3cm]{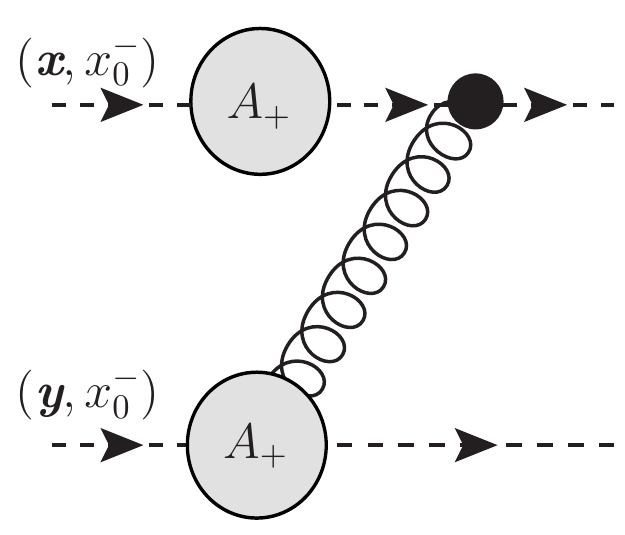}} + 
 \parbox{4cm}{\includegraphics[height=3cm]{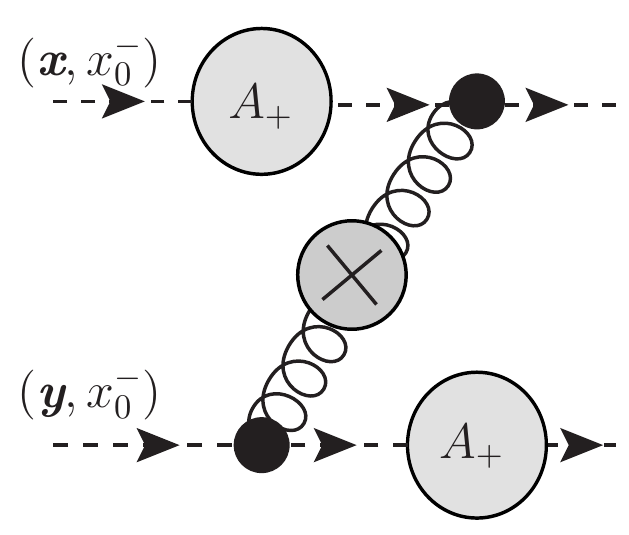}} +
 \parbox{4cm}{\includegraphics[height=3cm]{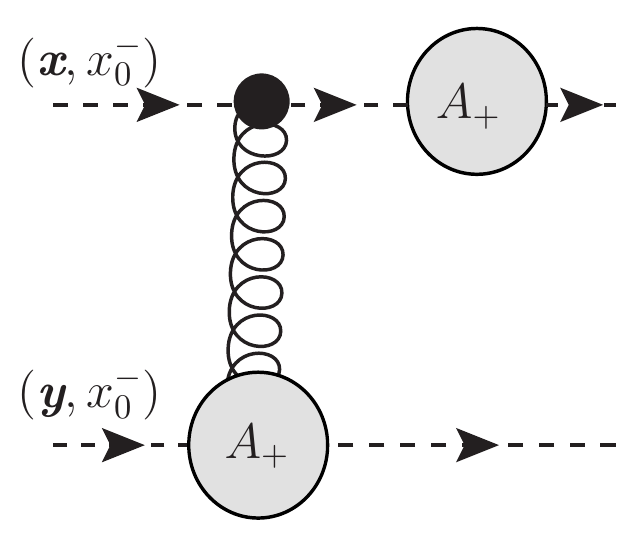}\,.}
\end{align}
Note that correlators of Wilson-lines are only infra-red finite, if
projected onto the color singlet. The general case of colored Wilson
lines is nevertheless of interest; in particular it allows to recover
the gluon Regge trajectory,  see
\cite{Caron-Huot:2013fea} for a detailed discussion. We therefore work in  $d =
4 + 2 \epsilon$ space-time dimensions, with  the vertices
Eq.~\eqref{eq:finally} and Eq.~\eqref{eq:quark_vertex} generalizing
trivially. We obtain
\begin{align}
  \label{eq:self2}
 & \parbox{4.5cm}{\center \includegraphics[height=2cm]{self2.pdf}}
 = (ig)^2 \! \int \frac{d^d p}{(2 \pi)^d} \!\int \frac{d^d r}{(2 \pi)^d}
  \frac{i}{-p^- - i \epsilon} \frac{i}{- r^- - i\epsilon}
   \frac{-i}{p^2 + i \epsilon} \frac{-i}{r^2 + i \epsilon}  \notag \\
&
2 \pi \delta(p^+ - r^+) \int d^{2 + 2 \epsilon} {\bm z}  e^{-i {\bm p }\cdot ({\bm x} - {\bm z})} 
 e^{-i {\bm r }\cdot ({\bm z} - {\bm x})}     t^b V({\bm x}) t^a  
\notag \\
& \hspace{6cm}
\cdot \left[ U^{ab}({\bm z}) - \delta^{ab} \right]-  \theta(-p^+)  \left[  [U^{ab}({\bm z})]^\dagger  - \delta^{ab} \right] \notag \\
& = 
\frac{g^2}{ \pi} \int_0^\infty \frac{dp^+}{p^+}    \int d^{2 + 2 \epsilon} {\bm z} \,  t^b V({\bm x}) t^a    \left[ U^{ab}({\bm z}) - \delta^{ab} \right]
\frac{\Gamma^2(1+\epsilon)}{(4\pi^{2 + 2 \epsilon})} \frac{({\bm x}- {\bm z})\cdot({\bm x}- {\bm z})}{[({\bm x}- {\bm z})^2]^{1+\epsilon}[({\bm x}- {\bm z})^2]^{1+\epsilon}}
\end{align}
The divergent integral over the plus-momenta provides the high-energy singularity which defines the kernel of the high energy evolution. The precise choice of the regulator is irrelevant for leading order accuracy. In the following  we chose  $\Lambda_{a,b} \to \infty$ and  a scale $s_0$   of the order of the transverse scale, also known as  the reggeization scale, to  regularize the integral as, 
\begin{align}
  \label{eq:reg2}
  \int \limits_{s_0/\Lambda^b}^{\Lambda^a} \frac{d p^+}{p^+} & =   \ln  \left(\frac{\Lambda_a \Lambda_b}{s_0} \right)\,.
\end{align}
To derive the high energy evolution of Wilson-lines, $\Lambda_a$ will
be the regulator of interest, since it limits the $p^+$ integral from
above.  With the $\overline{\text{MS}}$ strong coupling  constant in
$d=4 + 2 \epsilon$ dimensions
\begin{align}
  \label{eq:MSbaralphaS}
  \alpha_s & = \frac{g^2 \mu^{2\epsilon} \Gamma(1-\epsilon)}{(4 \pi)^{1 + \epsilon}},
\end{align}
we obtain
\begin{align}
  \label{eq:self2_fertig}
 \parbox{4.5cm}{\center \includegraphics[height=2cm]{self2.pdf}}  & = \ln  \left(\frac{\Lambda_a \Lambda_b}{s_0}  \right)\frac{\alpha_s}{\pi^2} \left(\frac{4}{\pi \mu^2} \right)^\epsilon
 \frac{\Gamma(1 + \epsilon)^2}{\Gamma(1-\epsilon)} \notag \\
&  \hspace{-2cm} \cdot 
 \int d^{2 + 2\epsilon} {\bm z}
\frac{({\bm x} - {\bm z})\cdot ({\bm x} - {\bm z}) }{[({\bm x} - {\bm z})^2]^{1 + \epsilon}[({\bm x} - {\bm z})^2]^{1 + \epsilon}} 
 t^b W({\bm x}) t^a \left[U^{ba}({\bm z}) - \delta^{ab} \right] \, .
\end{align}
We further have
\begin{align}
  \label{eq:self13}
 &  \parbox{4.5cm}{\center \includegraphics[height=2cm]{self1.pdf}} 
 +  \parbox{4.5cm}{\center \includegraphics[height=2cm]{self3.pdf}}
= \notag \\
& = \ln \left(\frac{\Lambda_a \Lambda_b}{s_0}  \right)\frac{\alpha_s \Gamma^2(1+\epsilon)}{2 \pi^2 \Gamma(1-\epsilon)} \left(\frac{4 }{\pi \mu^2} \right)^\epsilon  
 \int d^{2 + 2 \epsilon} {\bm z}  \frac{({\bm x} - {\bm z}) \cdot ({\bm x} - {\bm z})}{[({\bm x} - {\bm z})^2]^{1 + \epsilon}[({\bm x} - {\bm z})^2]^{1 + \epsilon}}
\notag \\
& \hspace{6cm}  \left[2 t^a W({\bm x}) t^a - t^a t^a W({\bm x}) - W({\bm x})t^a t^a \right] \, .
\end{align}
Combining both contributions one obtains
\begin{align}
  \label{eq:combineSELF}
 &  \ln \left( \frac{\Lambda_a \Lambda_b}{s_0} \right) \frac{\alpha_s \Gamma^2(1+\epsilon)}{2 \pi^2 \Gamma(1-\epsilon)} \left(\frac{4 }{\pi \mu^2} \right)^\epsilon  
 \int d^{2 + 2 \epsilon} {\bm z}  \frac{({\bm x} - {\bm z}) \cdot ({\bm x} - {\bm z})}{[({\bm x} - {\bm z})^2]^{1 + \epsilon}[({\bm x} - {\bm z})^2]^{1 + \epsilon}}
\notag \\
& \hspace{6cm}  \left[2  U^{ba}({\bm z}) t^b W({\bm x}) t^a - t^a t^a W({\bm x}) - W({\bm x})t^a t^a \right] \,.
\end{align}
The calculation for the interaction of two Wilson lines follows in
complete analogy:
\begin{align}
  \label{eq:4point_fertig}
 &  \parbox{4.5cm}{\center \includegraphics[height=3cm]{4point4.pdf}}  =  \ln \left(\frac{\Lambda_a \Lambda_b}{s_0}  \right)\frac{\alpha_s}{\pi^2} \left(\frac{4}{\pi \mu^2} \right)^\epsilon
 \frac{\Gamma(1 + \epsilon)^2}{\Gamma(1-\epsilon)}     \notag \\
&  \hspace{2cm}
 \cdot   \int d^{2 + 2\epsilon} {\bm z}
\frac{({\bm x} - {\bm z})\cdot ({\bm y} - {\bm z}) }{[({\bm x} - {\bm z})^2]^{1 + \epsilon}[({\bm y} - {\bm z})^2]^{1 + \epsilon}}  \quad
 t^b W({\bm x}) \otimes W({\bm y}) t^a \left[U^{ba}({\bm z}) - \delta^{ab} \right]\,,
\end{align}
and
\begin{align}
  \label{eq:self13}
 &  \parbox{4.5cm}{\center \includegraphics[height=3cm]{4point2.pdf}} 
 +  \parbox{4.5cm}{\center \includegraphics[height=3cm]{4point3.pdf}}
= \notag \\
& = \ln \left(\frac{\Lambda_a \Lambda_b}{s_0} \right) \frac{\alpha_s \Gamma^2(1+\epsilon)}{2 \pi^2 \Gamma(1-\epsilon)} \left(\frac{4 }{\pi \mu^2} \right)^\epsilon  
 \int d^{2 + 2 \epsilon} {\bm z}  \frac{({\bm x} - {\bm z}) \cdot ({\bm y} - {\bm z})}{[({\bm x} - {\bm z})^2]^{1 + \epsilon}[({\bm y} - {\bm z})^2]^{1 + \epsilon}}
\notag \\
& \hspace{0cm} 
 \left[ t^a W({\bm x}) \otimes W({\bm y}) t^a + 
   W({\bm x})t^a \otimes t^a W({\bm y})  - t^a  W({\bm x}) \otimes
  t^a  W({\bm y})  - W({\bm x})t^a \otimes W({\bm y})t^a \right] 
\end{align}
We then obtain for the complete correlator of 2 Wilson lines
\begin{align}
  \label{eq:20}
 &    \parbox{5cm}{\includegraphics[width=5cm]{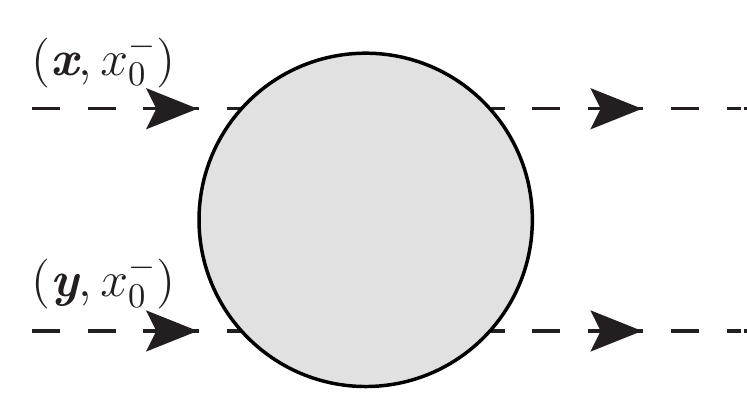}} = \ln \left(\frac{\Lambda_a \Lambda_b}{s_0} \right) \frac{\alpha_s \Gamma^2(1+\epsilon)}{2 \pi^2 \Gamma(1-\epsilon)} \left(\frac{4 }{\pi \mu^2} \right)^\epsilon  
 \int d^{2 + 2 \epsilon} {\bm z}
\notag \\
& 
\bigg\{ \frac{({\bm x} - {\bm z}) \cdot ({\bm x} - {\bm z})}{[({\bm x} - {\bm z})^2]^{1 + \epsilon}[({\bm x} - {\bm z})^2]^{1 + \epsilon}} \left[ 
2 U^{ab}({\bm z}) t^b W({\bm x}) t^a - t^a t^a W({\bm x}) - W({\bm x})t^at^a
\right]\otimes W({\bm y})\notag \\
&+ \frac{({\bm y} - {\bm z}) \cdot ({\bm y} - {\bm z})}{[({\bm y} - {\bm z})^2]^{1 + \epsilon}[({\bm y} - {\bm z})^2]^{1 + \epsilon}} \left[ 
2 U^{ab}({\bm z}) t^b W({\bm y}) t^a - t^a t^a W({\bm y}) - W({\bm y})t^at^a
\right]\otimes W({\bm x})\notag \\
&+
 \frac{({\bm x} - {\bm z}) \cdot ({\bm y} - {\bm z})}{[({\bm x} - {\bm z})^2]^{1 + \epsilon}[({\bm y} - {\bm z})^2]^{1 + \epsilon}} 
\bigg[-2 t^a W({\bm x}) \otimes t^a W({\bm y}) 
 -2  W({\bm x})t^a \otimes W({\bm y}) t^a \notag \\
& \hspace{4cm} +
2 U^{ab}({\bm z}) t^a W({\bm x}) \otimes W({\bm y})t^b + 
2 U^{ab}({\bm z}) t^a W({\bm y}) \otimes W({\bm x})t^b 
\bigg] \bigg\}\,.
\end{align}
Using the above result it is straightforward to obtain the high energy evolution of an ensemble of $n$ Wilson lines as 
\begin{align}
  \label{eq:2.6}
  - \Lambda_a \frac{d}{d \Lambda_a} \left[W({\bm x}_1) \otimes \ldots \otimes W({\bm x}_n) \right]
&=
\sum_{i,j = 1} H_{ij}  \left[W({\bm x}_1) \otimes \ldots \otimes W({\bm x}_n) \right]\,,
\end{align}
with the Balitsky-JIMWLK Hamiltonian
\begin{align}
  \label{eq:2.7}
  H_{ij} & = \frac{\alpha_s \Gamma^2(1+\epsilon)}{2 \pi^2 \Gamma(1-\epsilon)} \left(\frac{4 }{\pi \mu^2} \right)^\epsilon   \int d^{2+2\epsilon} {\bm z} 
\frac{({\bm x}_i - {\bm z}) \cdot ({\bm x}_j - {\bm z})}
{[({\bm x}_i - {\bm z})^2]^{1+\epsilon} [({\bm x}_j - {\bm z})^2]^{1+\epsilon}}
\notag \\
& \hspace{4cm}
\left[ 
T^a_{i, L} T^a_{j, L} + T^a_{i, R} T^a_{j, R} - U^{ab}({\bm z}) 
\left(T^a_{i, L} T^b_{j, R} + T^a_{j, L} T^b_{i, R}    \right)
\right].
\end{align}
In the presentation we followed here closely \cite{Caron-Huot:2013fea}
and define $T^a_{L,i}$ and $T^a_{R,j}$ as the group generators acting
to the left (L) or to the right (R) on the Wilson line $W({\bm x}_i)$,
\begin{align}
  \label{eq:2.5}
  T^a_{L,i} [W({\bm z}_i)]&  \equiv t^a W({\bm z}_i), &
 T^a_{R,i} [W({\bm z}_i)]&  \equiv W({\bm z}_i) t^a.
\end{align}

\section{Conclusion and Outlook}
\label{sec:conclusion-summary}

We investigated to which extent it is possible to obtain within
Lipatov's high energy effective action gluon and quark propagators,
which resum interaction with a strong (reggeized) gluon background
field, and whether the effective action allows to rederive
Balitsky-JIMWLK evolution. We found that both question can be answered
positively.To arrive at this result, we used a special parametrization
of the gluonic field, already proposed in \cite{Lipatov:1995pn}. This
parametrization allows both an expansion of the gluonic field around
the reggeized gluon field -- which is assumed to be strong -- and
provides consistent gauge transformation properties for the
parametrized gluonic field. Expanding the resulting effective
Lagrangian up to quadratic order in quantum fluctuations around the
strong reggeized gluon field, we obtain a new kind of
gluon-gluon-reggeized gluon vertex as well the usually
quark-quark-reggeized gluon vertex. Both vertices allow for a straightforward resummation of the reggeized gluon field to all orders into
Wilson lines. The resulting resummed gluon and quark propagators agree
for $v_-=0$ light-cone gauge with corresponding propagators which
include all order resummation of a gluonic background field in
light-cone gauge.  The latter are frequently employed in the
calculation of perturbative observables in the presence of high parton
densities, in particular within the Color Glass Condensate effective
theory. Finally we demonstrated that these propagators allow to
recover the complete (leading order) Balitsky-JIMWLK evolution
equation for Wilson lines from Lipatov's high energy effective action.
\\

Our results demonstrate that high energy factorization as formulated
within the Balitsky-JIMWLK evolution and high energy factorization as
formulated within Lipatov's high energy effective action are
equivalent. At the same time, Lipatov's high energy effective action
provides additional flexibility for actual calculations, since it
allows to adopt in a straightforward manner different gauges to
determine quantum fluctuations of the gluonic field. Moreover a
matching of results obtained within the BFKL-formalism and Lipatov's
high energy effective action on the one hand and light-front
perturbation theory and the Color-Glass-Condensate should be now
facilitated.  As an important side result we confirm the proposed
determination of the reggeized gluon from Balitsky-JIMWLK evolution
proposed in \cite{Caron-Huot:2013fea}, within the context of Lipatov's
high energy effective action.
\\

Future lines of research need to address the mentioned matching of NLO
results obtained within the two different frameworks as well as the
the explicit calculation of new NLO observables. Even though a number
of important NLO results have been obtained in the past for scattering
of a perturbative projectile on a dense target, see {\it
  i.e.}  \cite{nlocgc}, there is still a need to refine the available
tools for such calculations.  Another direction of research needs to
address the possible description of central production processes at
high parton densities as {\it i.e.}  required for the analysis of
nucleus-nucleus collisions and/or high multiplicity events. While the
current study is limited to the quasi-elastic region, such a program
requires the investigation of the corresponding effective action which
contains induced terms for both plus and minus reggeized gluon
fields. This is also related to the question whether such central
production terms can be formulated in a way which gives automatically
rise to the Balitsky-JIMWLK hierarchy. Related to this question is the
possible extension of Balitsky-JIMWLK evolution to exclusive
observables, generalizing already existing results
\cite{Hentschinski:2005er}.

\subsubsection*{Acknowledgments}
Conversations with Lev~N.~Lipatov, Jochen Bartels, Agustin Sabio Vera,
Grigorios Chachamis, Jose~D.Madrigal Martinez and Krzysztof Kutak on
the effective action and related topics as well as collaboration with
Alejandra Ayala, Jamal Jalilian-Marian and Maria-Elena Tejeda Yeomans
at an early stage of this project are gratefully acknowledged.

\appendix

\section{Multi-gluon exchange within the high energy effective
  action}
\label{sec:multi-gluon-exchange}

\begin{figure}[th]
  \centering
 \parbox{2.7cm}{ \includegraphics[width=2.5cm]{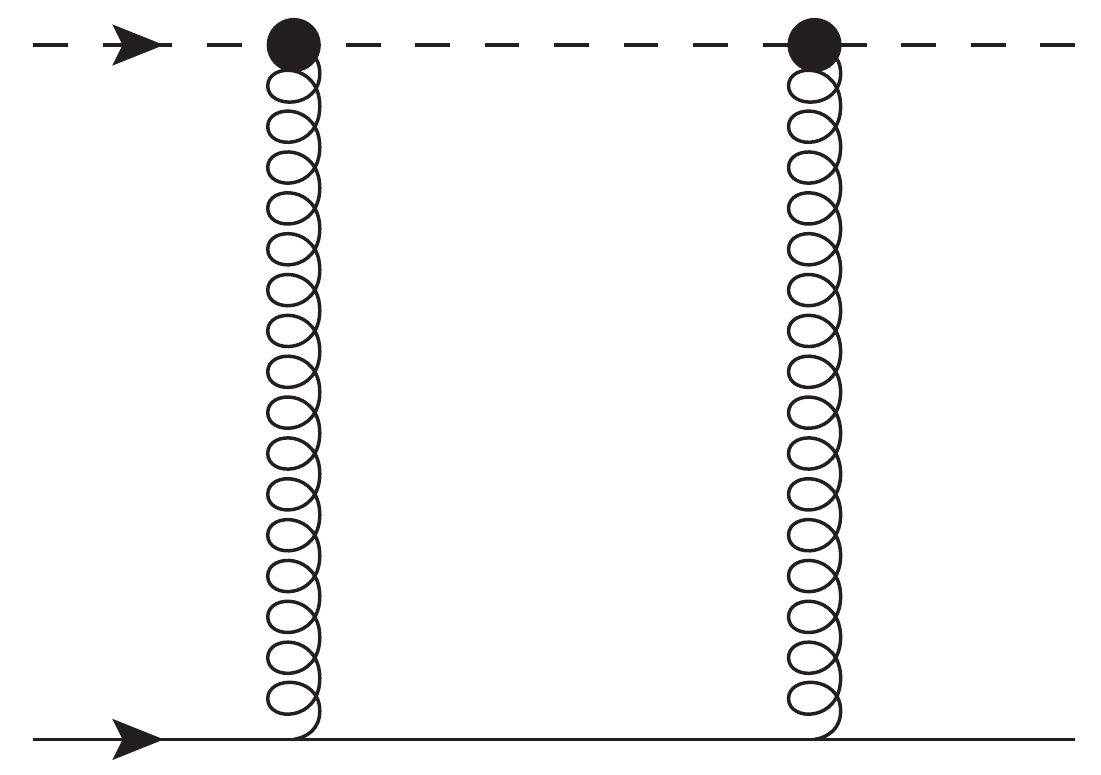}}
\parbox{2.7cm}{ \includegraphics[width=2.5cm]{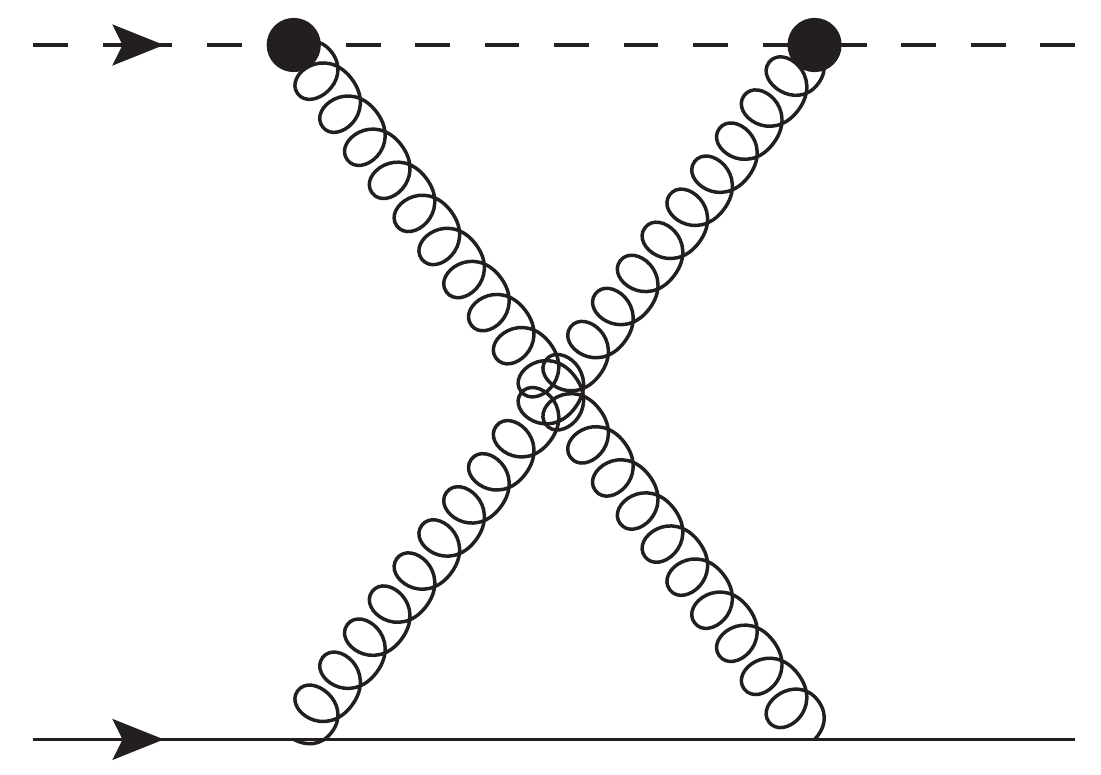}}
$\qquad$
\parbox{2.7cm}{ \includegraphics[width=2.5cm]{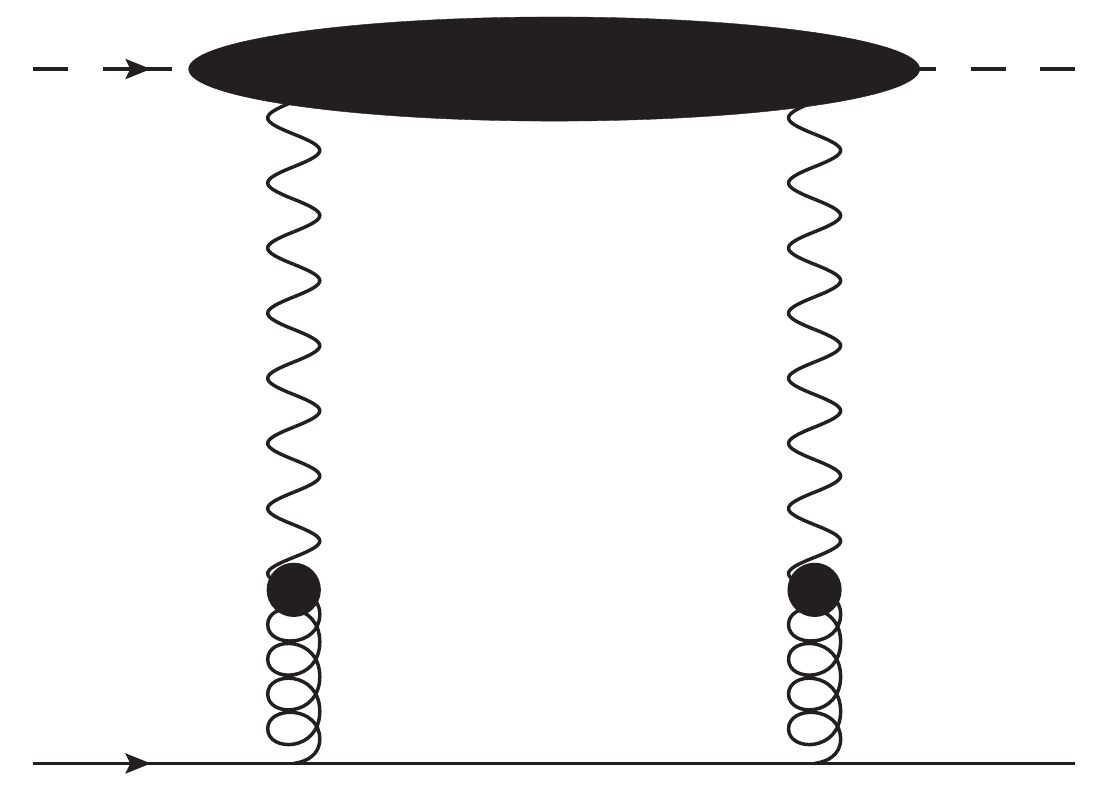}}
\parbox{2.7cm}{ \includegraphics[width=2.5cm]{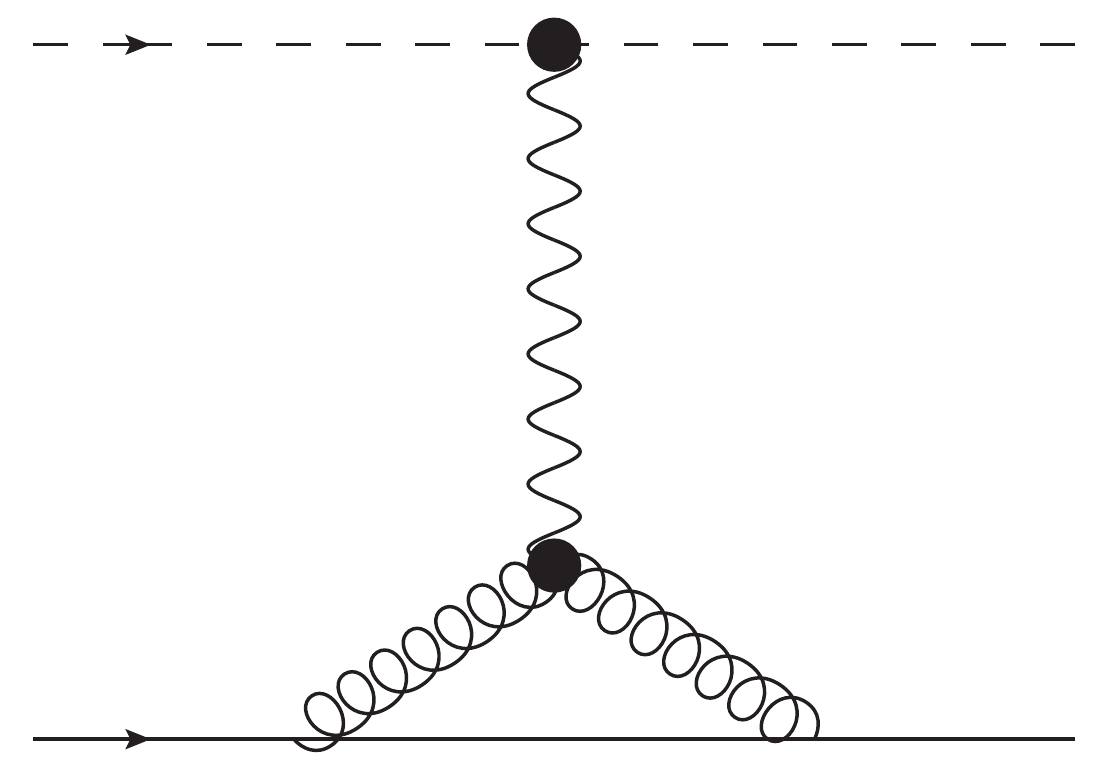}}
\parbox{2.7cm}{ \includegraphics[width=2.5cm]{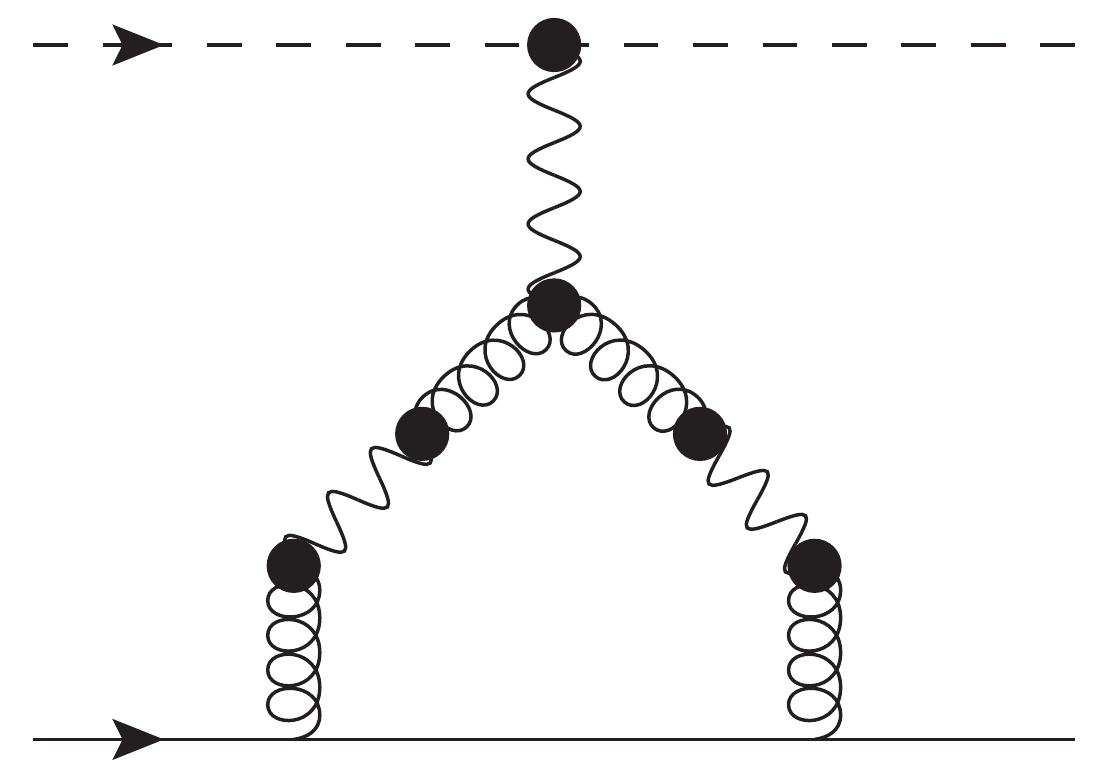}}\\
 \parbox{2.7cm}{\center (a)}
\parbox{2.7cm}{\center (b)}
$\qquad$
\parbox{2.7cm}{\center (c)}
\parbox{2.7cm}{\center (d)}
\parbox{2.7cm}{\center (e)}
  \caption{\small {\it Left:} 2 gluon exchange within QCD. {\it} Right: The corresponding decomposition within the high energy effective action in symmetric (2 reggeized gluon exchange) and anti-symmetric contribution}
  \label{fig:2gluons}
\end{figure}

We consider in the following the interaction of a Wilson line in the
fundamental representation with a color current, where the interaction
is mediated through the exchange of reggeized gluons. To embed the Wilson line into
a physical process (and to take the regarding high energy limit), one
can for instance use the vertex Eq.~\eqref{eq:quark_vertex}, and
combine it with corresponding quark spinors; this relates then the
following discussion to scattering of a quark on a color current.  For
definiteness we take for the color current on which the Wilson is
scattering a quark. The following result does not depend on those
details. We are further only interested in $t$-channel gluon exchange
of (high energy) gluons between the Wilson line and the color current;
couplings of the reggeized gluon to the quark take therefore place
through the QCD quark-gluon vertex as well as induced vertices Fig.~\ref{fig:3}. \\

Starting with two gluon exchange as the first non-trivial contribution
we have within conventional QCD the two diagrams depicted in
Fig.~\ref{fig:2gluons}.a-b, while the two relevant contributions
within the high energy effective action are given in
Fig.~\ref{fig:2gluons}.c-e. The black blob denotes the various
couplings of the reggeized gluon to the Wilson line. For two reggeized
gluons one has
\begin{align}
  \label{eq:ip2}
  \parbox{3cm}{\includegraphics[width=3cm]{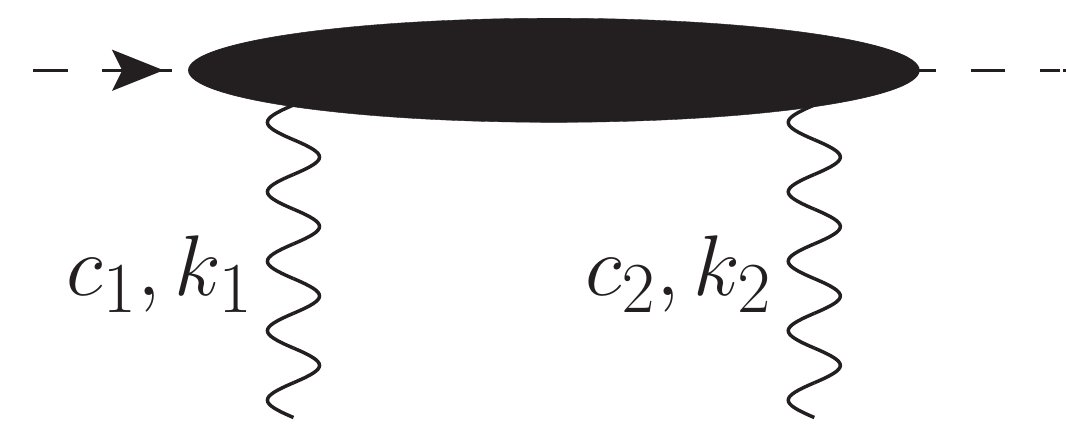}} &  = 
  \parbox{3cm}{\includegraphics[width=3cm]{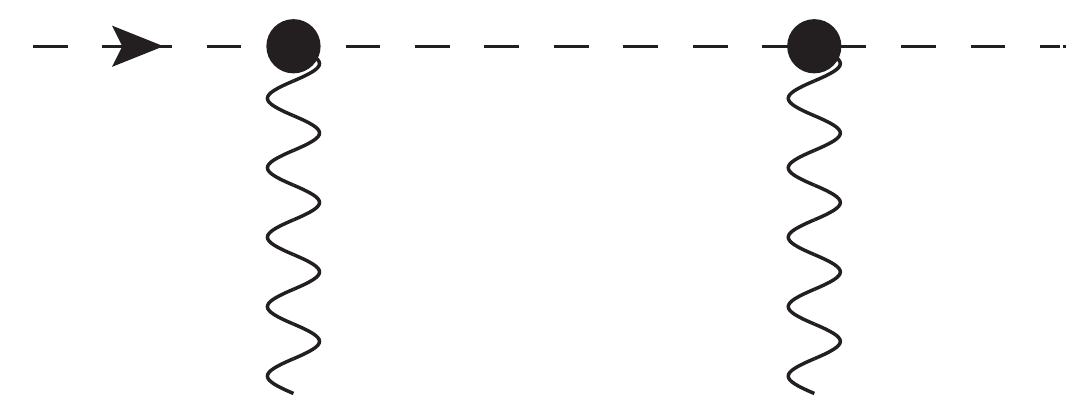}} +  
  \parbox{3cm}{\includegraphics[width=3cm]{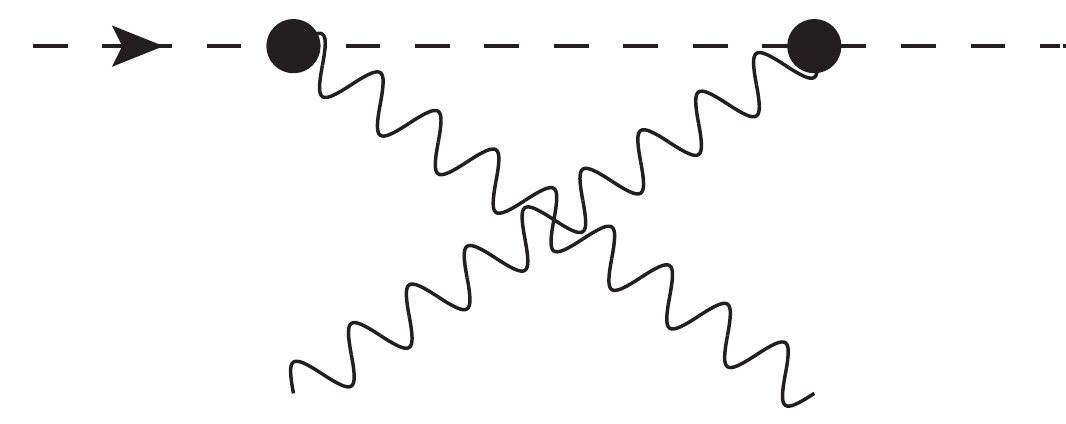}} 
\notag \\
& = 
ig t^{c_2} \frac{i}{k^-_1 + i\epsilon} ig t^{c_1} 
+   
ig t^{c_1} \frac{i}{k^-_2 + i\epsilon} ig t^{c_2}\, .
\end{align}
Due to high-energy kinematics, the loop integral in the diagram with
two reggeized gluon exchange of Fig.~\ref{fig:2gluons}.c factorizes. It
is therefore possible to associate the integration over minus momentum
directly with the Wilson line:
\begin{align}
  \label{eq:ip2int}
  \int \frac{dk_1^-}{2 \pi} \int \frac{dk_2^-}{2 \pi} \,  2\pi \delta(k_1^- + k_2^-) \quad 
 \parbox{3cm}{\includegraphics[width=3cm]{IPregg21.pdf}}
&=
(ig)^2 \cdot \frac{1}{2} \left(t^{c_1}t^{c_2} +  t^{c_2}t^{c_1}
  \right) 
\notag \\
&= (ig)^2 S_2(12)\,.
\end{align}
In the above we used a short-hand notation, introduced in \cite{Hentschinski:2011xg},
\begin{align}
  \label{eq:notation}
  [i,j] & \equiv [t^{c_i}, t^{c_j}] & S_n(1\ldots n) \equiv \frac{1}{n!} \sum_{i_1,\ldots, i_n} t^{c_{i_1}} \cdots   t^{c_{i_n}}
\end{align}
 where in the second term the sum is taken over all permutations of the numbers $1, \ldots, n$. Using this notation, a possible decomposition of a color tensor with two adjoint color indices is given by the following basis,
\begin{align}
  \label{eq:decomp2}
  [1,2], &&& S_2(12) \,.
\end{align}
In \cite{Hentschinski:2011xg}, this decomposition has been used to
construct the the pole prescription for induced vertices, by
projecting out the anti-symmetric sector  of the complete color
structure of a Wilson line. Using this pole prescription and
associating the integration over minus momentum similar with the 1
reggeized gluon to 2 reggeized gluon splitting, similar to
Eq.~\eqref{eq:ip2int}, it is then straightforward to demonstrate that
diagrams such as Fig.~\ref{fig:2gluons}.e vanishes. We note that this
holds for all splittings of a single reggeized gluons into $n$
reggeized gluons at tree-level, {\it i.e.} such splittings are
generally absent within this particular pole prescription\footnote{We
  note that a prescription different from the one of
  \cite{Hentschinski:2011xg} has been used in \cite{Braun:2017qij}. We
  point out the possibility that the arguments presented here may not
  hold for this particular prescription.} after integration over
corresponding light-cone momenta. The only diagrams left are therefore
Fig.~\ref{fig:2gluons}.c and Fig.~\ref{fig:2gluons}.d, where the
induced vertex associated with Fig.~\ref{fig:2gluons}.d, carries the
color tensor $[1,2]$, providing therefore the anti-symmetric
contribution missing in Eq.~\eqref{eq:ip2int}. For an explicit
decomposition of diagrams such as Fig.~\ref{fig:2gluons}.a and
Fig.~\ref{fig:2gluons}.b, we refer the interested reader to
\cite{Hentschinski:2009zz,Hentschinski:2011xg}. \\

\begin{figure}[th]
  \centering
  \parbox{2.5cm}{ \includegraphics[width=2.5cm]{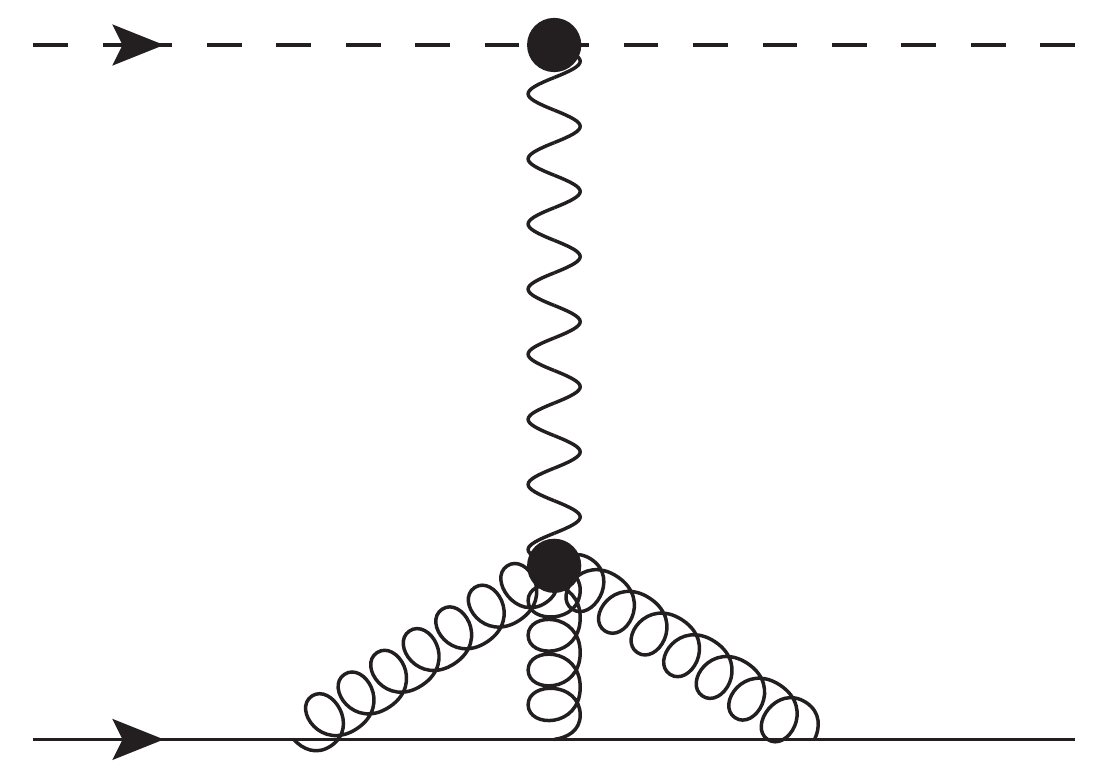}}  $\qquad$
  \parbox{2.5cm}{ \includegraphics[width=2.5cm]{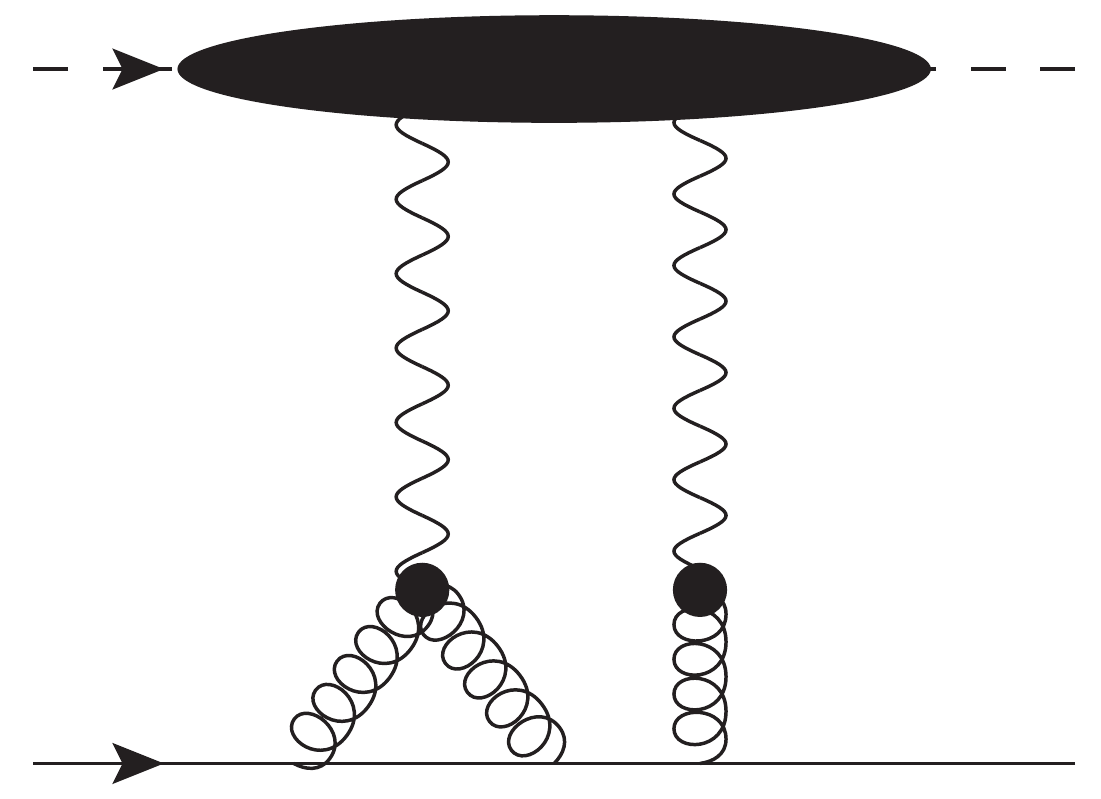}}
  \parbox{2.5cm}{ \includegraphics[width=2.5cm]{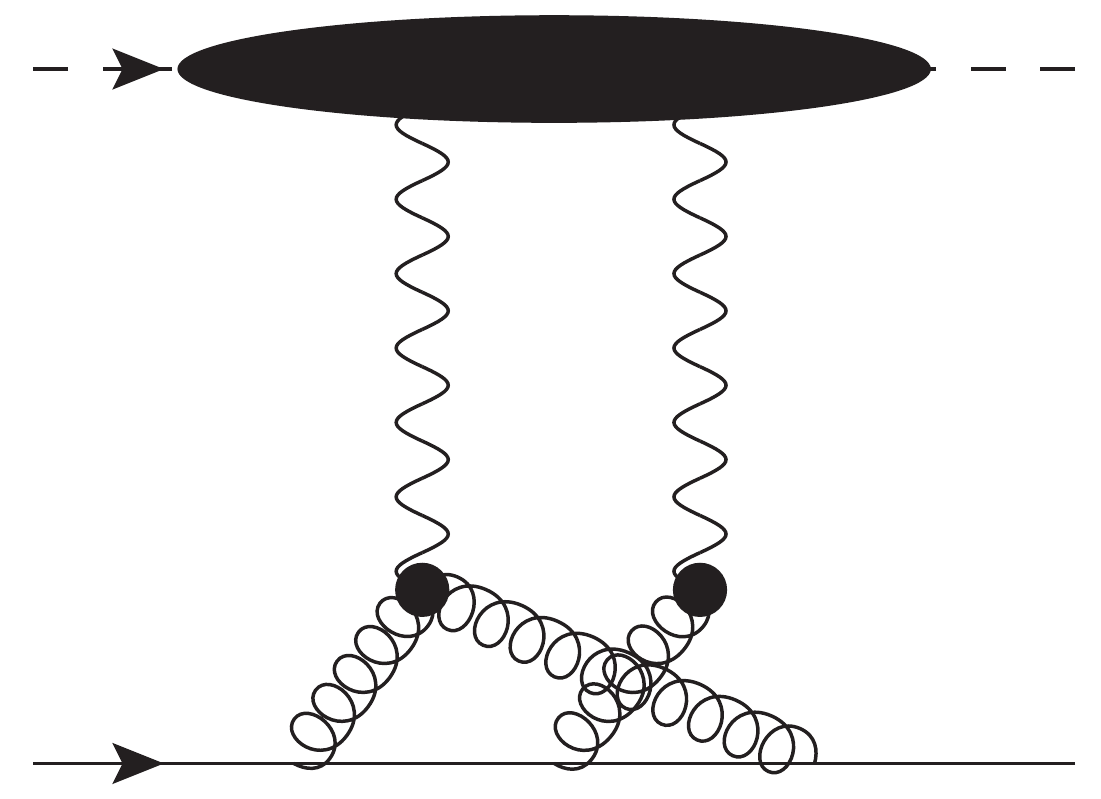}}
  \parbox{2.5cm}{ \includegraphics[width=2.5cm]{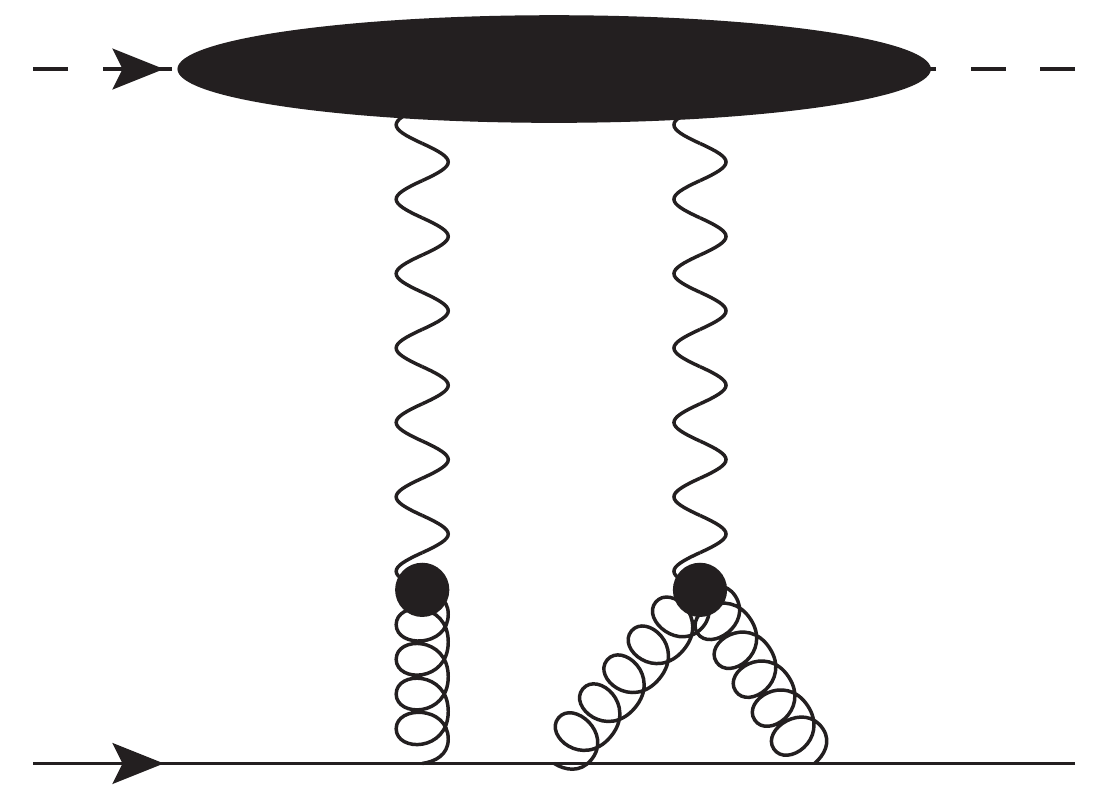}} $\qquad $
  \parbox{2.5cm}{ \includegraphics[width=2.5cm]{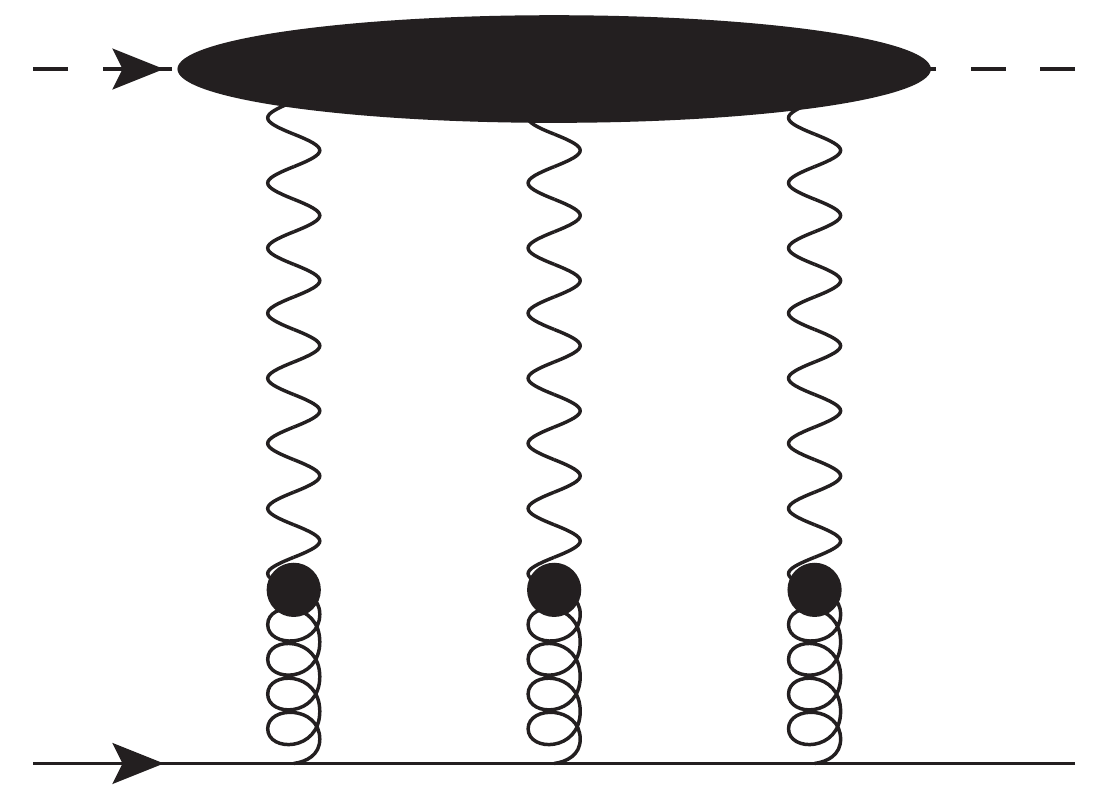}} \\

\parbox{2.5cm}{\center (a)}  $\qquad$
\parbox{2.5cm}{\center (b)} 
\parbox{2.5cm}{\center (c)} 
\parbox{2.5cm}{\center (d)}  $\qquad$
\parbox{2.5cm}{\center (e)} 
 \caption{\small Three gluon exchange within the high energy effective action. 
{\it Left:} The anti-symmetric contribution. {\it Center:} The contribution with mixed symmetry. {\it Right:} The symmetric contribution. }
  \label{fig:3reggeized}
\end{figure}
The corresponding symmetry decomposition for three adjoint color
indices is provided by the following six tensors: 
\begin{align}
  \label{eq:decomp3}
  [[3,1],2], && [[3,2],1], && S_2([1,2]3), && S_2([1,3]2),  && S_2([2,3]1), && S_3(123)\,.
\end{align}
With 
\begin{align}
  \label{eq:ip2}
  \parbox{2.5cm}{\includegraphics[width=2.5cm]{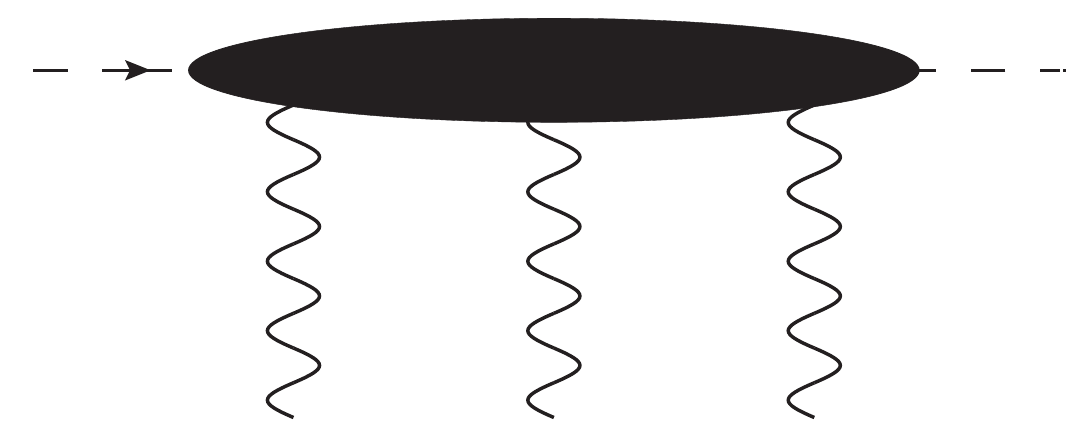}} &  = 
  \parbox{2.5cm}{\includegraphics[width=2.5cm]{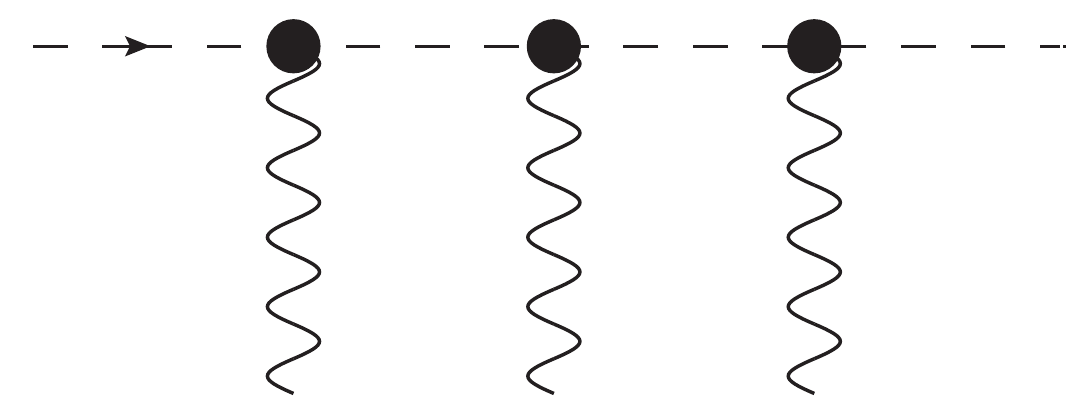}} +  
  \parbox{2.5cm}{\includegraphics[width=2.5cm]{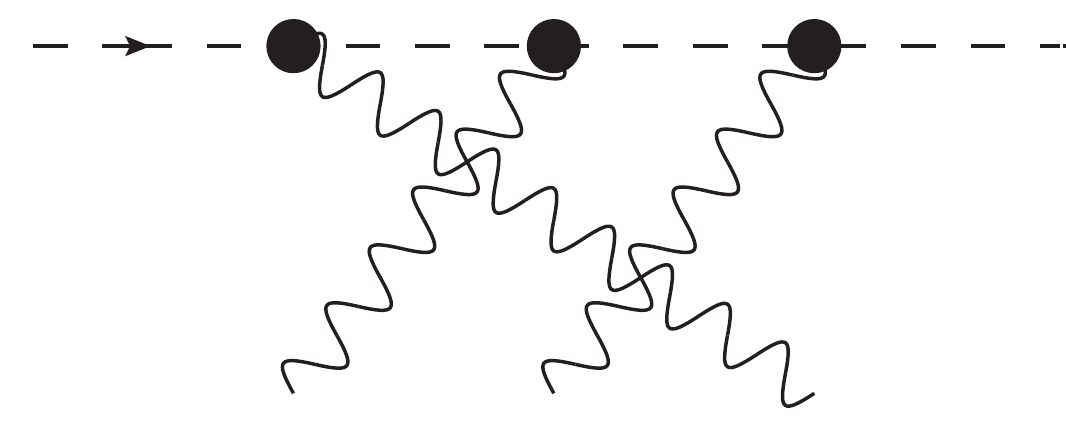}} +
  \parbox{2.5cm}{\includegraphics[width=2.5cm]{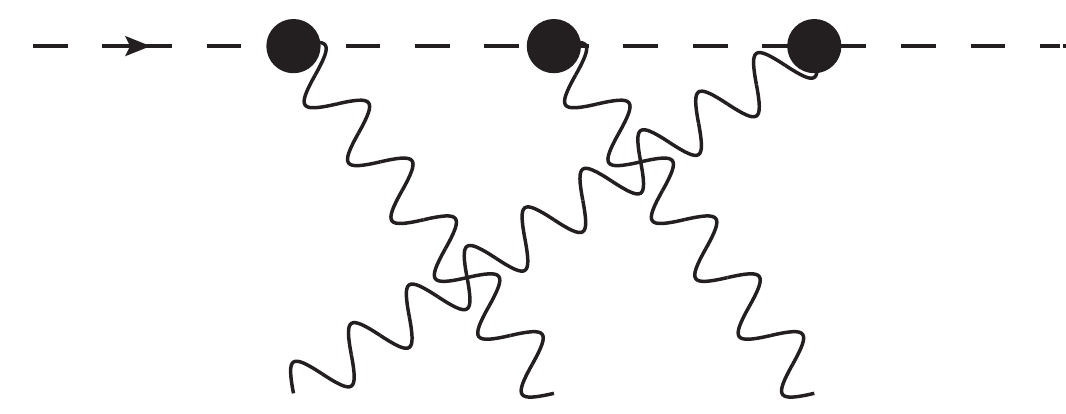}} 
\notag \\
&+  
  \parbox{2.5cm}{\includegraphics[width=2.5cm]{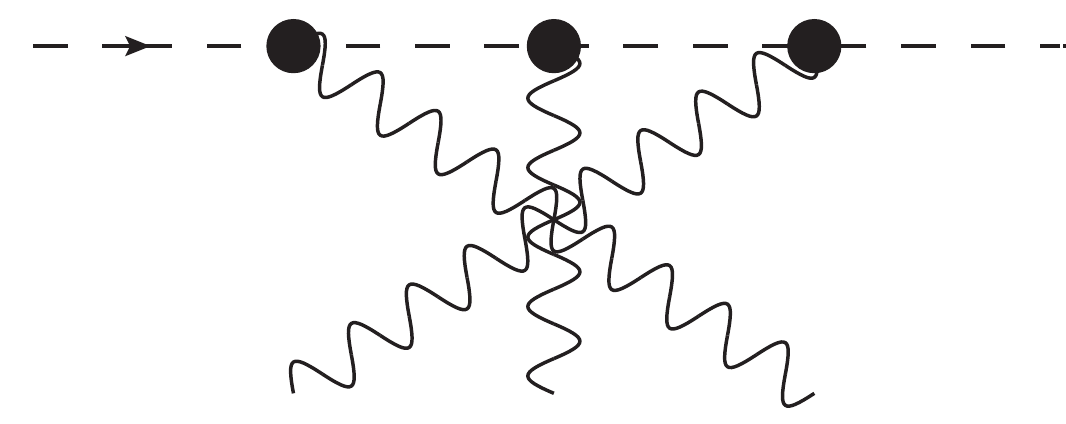}} +
  \parbox{2.5cm}{\includegraphics[width=2.5cm]{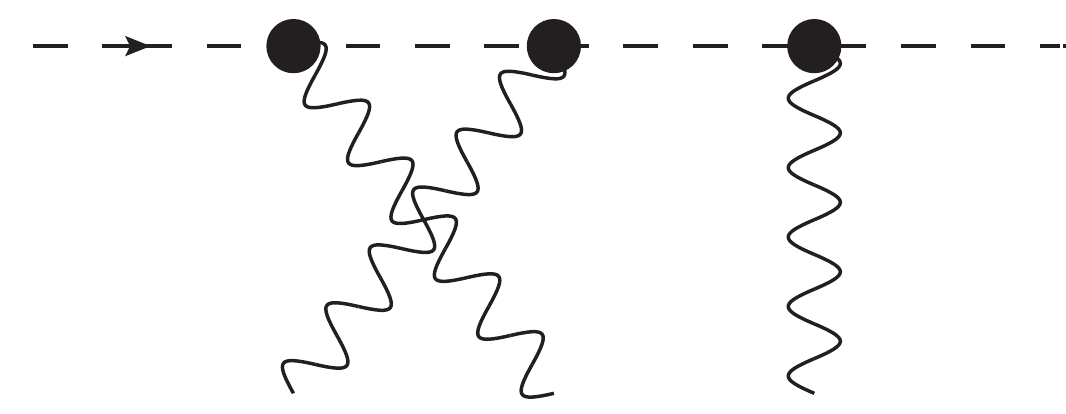}} +  
  \parbox{2.5cm}{\includegraphics[width=2.5cm]{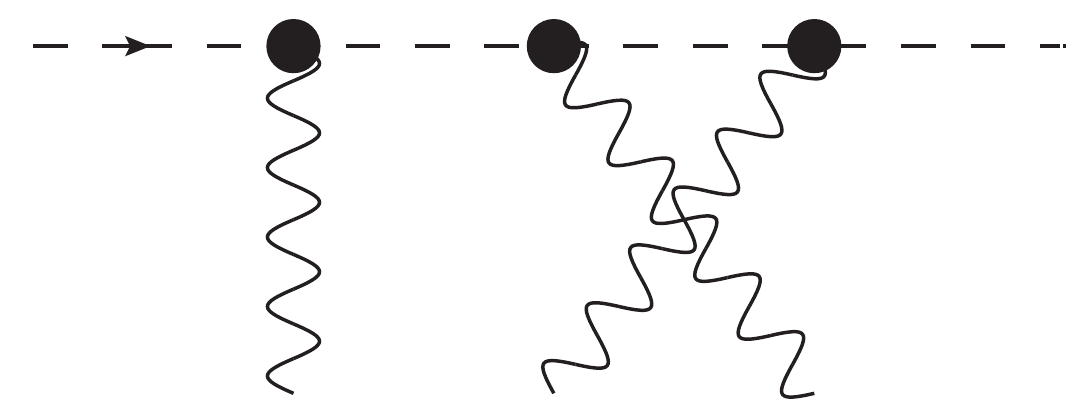}} \,.
\end{align}
It is straightforward to demonstrate that
\begin{align}
  \label{eq:int3gluon}
  \int \frac{d k_1^-}{2 \pi}     
 \int \frac{d k_2^-}{2 \pi}     
\int \frac{d k_3^-}{2 \pi}   2 \pi \delta(k_1^- + k_2^- + k_3^-)
 \parbox{2.5cm}{\includegraphics[width=2.5cm]{IPregg30.pdf}} = (ig)^3 S_3(123)\,.
\end{align}
We therefore find that Fig.~\ref{fig:3reggeized}.a represents the
color tensor $[[3,1],2]$ and $[[3,2],1]$ through the color tensors
contained in the induced vertex Fig.~\ref{fig:3}.
Fig.~\ref{fig:3reggeized}.b-d provide the color tensors $S_2([1,2]3)$,
$S_2([1,3]2)$, and $S_2([2,3]1)$, through the combination of the
symmetric 2 reggeized gluon state with the induced vertex. Finally
Fig.~\ref{fig:3reggeized}.e provides the color tensor $ S_3(123)$. For
the explicit construction of the Wilson line with three gluons
decomposed into the above color tensors, we refer the interested
reader again to \cite{Hentschinski:2009zz,Hentschinski:2011xg}, where
furthermore some details on the four gluon exchange can be found. The
general picture should be nevertheless already clear at this stage:
even though the color tensor associated with $n$ reggeized gluons
coupled to a Wilson line is automatically symmetric, the high energy
effective action is capable to construct the complete color structure
provided by path ordered gluons making use of the additional induced
vertices. The latter provide the necessary anti-symmetric color
tensors as well as corresponding terms of mixed symmetry if combined
with multiple reggeized gluon exchange.

\section{Quantum fluctuations of Wilson line}
\label{sec:quant-fluct-wils}

In the following we provide further details on the derivation of the Feynman rules  for the calculation of Wilson lines.  The propagator without fluctuations is easily obtained from the action Eq.~\eqref{eq:Lagrangian_scalar_shift}. In particular
\begin{align}
  \label{eq:propagator_W}
 \left \langle \infty \left |  \frac{1}{1 + \frac{g}{\partial_+ +  \epsilon} A_+} \frac{1}{\partial_+ + \epsilon} \right | - \infty  \right \rangle & = \mathrm{P} \exp \left(\frac{-g}{2} \int_{-\infty}^{\infty} d z^+  A_+ \right) = W[A_+]({\bm z}, x_0^-).
\end{align}
To include fluctuations, we first consider the case
$A_+ \to A_+ + [A_+, \frac{g}{\partial_-} v_-]$.  Since  $\partial_- A_+ = 0$, the operator $1/\partial_-$ does not act on the $A_+$-fields.
We therefore consider a shift of the form 
$A_+(x) \to  A_+(x) + [A_+(x), w(x)] = A_+(x) + [A_+(x), w(0, {\bm x},
x^-)]$ with $w(x) = \frac{g}{\partial_-} v_-(x)$
and where we used the delta function implicitly contained in $A_+$ to set
$x^+ = 0$ in the fluctuation $w(x)$. Expanding to linear order in $w$,
\begin{align}
  \label{eq:expandW}
  \left(A_+ +  [A_+(x), w(x)] \right)^n & =  A_+^n +  \sum_{i=0}^{n-1}   \left( A_+^{i+1} w  A_+^{n-i-1} -  A_+^{i} w   A_+^{n-i} \right) + \mathcal{O}(w^2) 
\notag \\
& =  A_+^n  +  \sum_{i=1}^{n}  A_+^{i} w  A_+^{n-i} 
-   \sum_{i=0}^{n-1}  A_+^{i} w  A_+^{n-i}   + \mathcal{O}(w^2)  \notag \\
& =  A_+^n  +  A_+^n  w - w  A_+^n   + \mathcal{O}(w^2).
\end{align}
one finds
\begin{align}
  \label{eq:f;d}
W\left[A^+ + [A^+, w]\right]( {\bm x}, x^-) 
 = 
  W[A^+]  +   W[A^+]  \cdot w(x) -  w(x) \cdot    W[A^+]  + \mathcal{O}(w^2)
\end{align}
where  $ w (x) =  w (x^+ = 0, {\bm x}, x^-)$. The fluctuation $A_+ \to A_+ + [A_+, \frac{g}{\partial_-} v_-]$ leads therefore to
\begin{align}
  \label{eq:commutatorstuff_done}
\frac{g}{\partial_- }   \left[ W[A_+]({\bm x}, x_0^-), v_-(x) \right]
& = 
 \frac{g}{2} \int_{-\infty}^{x_0^-} dx^-  \left[ W[A_+]({\bm x}, x_0^-), v_-(x) \right],
\end{align}
which translates directly into the Feynman rule Fig.~\ref{fig:feynman_wilson}.c.
 The second type of fluctuations requires a shift of the form,
\begin{align}
  \label{eq:17}
  V_+(x) &=  A_+(x) + v_+(x),
\end{align}
where $v_+$ does not have delta-like support.  One finds to linear
order in the fluctuations $v_+(x)$,
\begin{align}
  \label{eq:solutionshift}
 &  W[A + v](x)\bigg|_{x^+ = \infty}    = W[A](x)\bigg|_{x^+ = \infty} 
\notag \\
&  + 
 \sum_{n=0}^\infty \left(\frac{-g}{2} \right)^n \prod_{i=1}^n
  \int\! d x_i^+  \sum_{j=1}^n A_+(x_1) \ldots A_+(x_{j-1}) v_+(x_{j}) A_+(x_{j+1}) \ldots A_+(x_{n}) 
\notag \\
& \hspace{2cm}
 \theta(x_1^+ - x_2^+) \ldots \theta(x^+_{j-1} - x_j^+) 
 \theta(x^+_j - x_{j+1}^+) \ldots \theta(x^+_{n-1} - x_n^+) + \mathcal{O}(v_+^2).
\end{align}
$A_+(x) \sim \delta(x^+-x^+_0)$ sets now
$ \theta(x^+_{j-1} - x^+_j) \theta(x^+_{j} - x^+_{j+1}) \to \theta
(x^+_0- x^+_j) \theta(x^+_{j} - x^+_0)$.
The integral over $x_j^+$ has therefore zero support and yields zero
result.  The only contributions which remain are $j=1$ and $j=n$, {\it
  i.e.} the cases where the $v_+$ is placed as the first or the last
term. For term with $m$ fluctuations one therefore finds
\begin{align}
  \label{eq:solutionshift}
& \left(\frac{-g}{2} \right)^m  \sum_{n=0}^m \prod_{i=1}^n \prod_{j=n+1}^m
  \int_{x^+_0}^\infty d x_i^+  \int_{-\infty}^{x^+_0} d x_j^+    v_+(x_1) \ldots v_+(x_n) W[A_+]  v_+(x_{n+1}) \ldots v_+(x_m) 
\notag \\
& \hspace{2cm}
 \theta(x_1^+ - x_2^+) \ldots \theta(x^+_{n-1} - x^+_n) 
 \theta(x^+_{n+1} - x_{n+1}^+) \ldots \theta(x^+_{m-1} - x_m^+)  \,.
\end{align}
Fluctuations $A_+ \to A_+ + v_+$ are therefore taken into account
through a Wilson-line gluon vertex, see Fig.~\ref{fig:feynman_wilson}
which can be inserted only before or after the $A_+$ fields to the
Wilson line.


\begin{thebibliography}{99}

\bibitem{Fadin:1975cb} 
  V.~S.~Fadin, E.~A.~Kuraev and L.~N.~Lipatov,
  Phys.\ Lett.\  {\bf 60B}, 50 (1975);
  Sov.\ Phys.\ JETP {\bf 44}, 443 (1976)
  [Zh.\ Eksp.\ Teor.\ Fiz.\  {\bf 71}, 840 (1976)];
  Sov.\ Phys.\ JETP {\bf 45}, 199 (1977)
  [Zh.\ Eksp.\ Teor.\ Fiz.\  {\bf 72}, 377 (1977)].

\bibitem{Balitsky:1978ic}
I.~I. Balitsky and L.~N. Lipatov, ``{The Pomeranchuk Singularity in Quantum
  Chromodynamics},'' {\em Sov.J.Nucl.Phys.} {\bf 28} (1978) 822--829.

\bibitem{Fadin:1998py}
  V.~S.~Fadin, L.~N.~Lipatov, Phys.\ Lett.\  B {\bf 429} (1998) 127;
  M.~Ciafaloni, G.~Camici, Phys.\ Lett.\  B {\bf 430} (1998) 349.

\bibitem{Bautista:2016xnp} 
M.~Hentschinski, A.~Sabio Vera and C.~Salas,
  Phys.\ Rev.\ Lett.\  {\bf 110} (2013) no.4,  041601
  [arXiv:1209.1353 [hep-ph]]; 
  Phys.\ Rev.\ D {\bf 87}, no. 7, 076005 (2013)
  [arXiv:1301.5283 [hep-ph]]; 
  I.~Bautista, A.~Fernandez Tellez and M.~Hentschinski,
  Phys.\ Rev.\ D {\bf 94}, no. 5, 054002 (2016)
  [arXiv:1607.05203 [hep-ph]].


\bibitem{gijv} 
  F.~Gelis, E.~Iancu, J.~Jalilian-Marian and R.~Venugopalan,
  Ann.\ Rev.\ Nucl.\ Part.\ Sci.\  {\bf 60}, 463 (2010)
  [arXiv:1002.0333 [hep-ph]].

\bibitem{Balitsky:1995ub} 
  I.~Balitsky,
  Nucl.\ Phys.\ B {\bf 463}, 99 (1996)
  [hep-ph/9509348].


\bibitem{jimwlk1}
J.~Jalilian-Marian, A.~Kovner, L.~D.~McLerran and H.~Weigert,
Phys.\ Rev.\  D {\bf 55}, 5414 (1997); 
 J.~Jalilian-Marian, A.~Kovner, A.~Leonidov and H.~Weigert,
  Phys.\ Rev.\ D {\bf 59} (1998) 014014
  [hep-ph/9706377];
Phys.\ Rev.\  D {\bf 59}, 034007 (1999).  
J.~Jalilian-Marian, A.~Kovner,  and H.~Weigert
Phys.\ Rev.\  D {\bf 59}, 014015 (1998).

\bibitem{jimwlk2}
J.~Jalilian-Marian, A.~Kovner, A.~Leonidov and H.~Weigert,
Nucl.\ Phys.\  B {\bf 504}, 415 (1997), 


\bibitem{jimwlk6}
A.~Kovner, J.~G.~Milhano and H.~Weigert,
Phys.\ Rev.\  D {\bf 62}, 114005 (2000); 
A.~Kovner and J.~G.~Milhano,
Phys.\ Rev.\  D {\bf 61}, 014012 (1999).
\bibitem{jimwlk8}
E.~Iancu, A.~Leonidov and L.~D.~McLerran,
Nucl.\ Phys.\  A {\bf 692}, 583 (2001); 
Phys.\ Lett.\  B {\bf 510}, 133 (2001);
E.~Ferreiro, E.~Iancu, A.~Leonidov and L.~McLerran,
Nucl.\ Phys.\  A {\bf 703}, 489 (2002).




\bibitem{Lipatov:1995pn}
  L.~N.~Lipatov,
  Nucl.\ Phys.\  {\bf B452 } (1995)  369-400   
  [arXiv:hep-ph/9502308 [hep-ph]].
\bibitem{Lipatov:1996ts}
  L.~N.~Lipatov,
  Phys.\ Rept.\  {\bf 286 } (1997)  131-198 
  [hep-ph/9610276]. 


\bibitem{quarkjet} 
  M.~Hentschinski and A.~Sabio~Vera,
  Phys.\ Rev.\ D {\bf 85}, 056006 (2012)
  [arXiv:1110.6741 [hep-ph]].
\bibitem{gluonjet} 
  G.~Chachamis, M.~Hentschinski, J.~D.~Madrigal and A.~Sabio~Vera, Phys.~Rev.~D {\bf 87} (2013) 076009
  [arXiv:1212.4992 [hep-ph]].


\bibitem{Hentschinski:2014esa}
  M.~Hentschinski, J.~D.~M.~Mart\'inez, B.~Murdaca and A.~Sabio Vera,
  Nucl.\ Phys.\ B {\bf 889} (2014) 549
  [arXiv:1409.6704 [hep-ph]];
  Nucl.\ Phys.\ B {\bf 887} (2014) 309
  [arXiv:1406.5625 [hep-ph]];
  Phys.\ Lett.\ B {\bf 735} (2014) 168
  [arXiv:1404.2937 [hep-ph]].


\bibitem{traject}
G.~Chachamis, M.~Hentschinski, J.~D.~Madrigal and A.~Sabio Vera,
  Nucl.\ Phys.\ B {\bf 876}, 453 (2013)
  [arXiv:1307.2591 [hep-ph]];
  Nucl.\ Phys.\ B {\bf 861}, 133 (2012)
  [arXiv:1202.0649 [hep-ph]].

\bibitem{Bartels:2012sw} 
  J.~Bartels, V.~S.~Fadin, L.~N.~Lipatov and G.~P.~Vacca,
  Nucl.\ Phys.\ B {\bf 867}, 827 (2013)
  doi:10.1016/j.nuclphysb.2012.10.024
  [arXiv:1210.0797 [hep-ph]].


\bibitem{review} 
  G.~Chachamis, M.~Hentschinski, J.~D.~Madrigal and A.~Sabio~Vera,
  Phys.\ Part.\ Nucl. {\bf 45}, 788 (2014) [arXiv:1211.2050 [hep-ph]].



\bibitem{Nefedov:2017qzc}
  M.~Nefedov and V.~Saleev,
  Mod.\ Phys.\ Lett.\ A {\bf 32} (2017) no.40,  1750207
  [arXiv:1709.06246 [hep-th]].

\bibitem{Braun:2017qij}
  M.~A.~Braun and M.~Y.~Salykin,
  Eur.\ Phys.\ J.\ C {\bf 77} (2017) no.7,  498
  [arXiv:1702.04796 [hep-ph]];
  M.~A.~Braun, M.~B.~Vyazovsky, S.~S.~Pozdnyakov and M.~Y.~Salykin,
  Bull.\ Russ.\ Acad.\ Sci.\ Phys.\  {\bf 80} (2016) no.8,  959
   [Izv.\ Ross.\ Akad.\ Nauk Ser.\ Fiz.\  {\bf 80} (2016) no.8,  1047];
  M.~A.~Braun and M.~I.~Vyazovsky,
  Phys.\ Rev.\ D {\bf 93} (2016) no.6,  065026
  [arXiv:1601.03469 [hep-ph]];
  M.~A.~Braun, S.~S.~Pozdnyakov, M.~Y.~Salykin and M.~I.~Vyazovsky,
  Eur.\ Phys.\ J.\ C {\bf 75} (2015) no.5,  222
  [arXiv:1502.03152 [hep-ph]];
  M.~A.~Braun, S.~S.~Pozdnyakov, M.~Y.~Salykin and M.~I.~Vyazovsky,
  Eur.\ Phys.\ J.\ C {\bf 74} (2014) no.8,  2989
  [arXiv:1402.4786 [hep-ph]];
  M.~A.~Braun, L.~N.~Lipatov, M.~Y.~Salykin and M.~I.~Vyazovsky,
  Eur.\ Phys.\ J.\ C {\bf 71} (2011) 1639
  [arXiv:1103.3618 [hep-ph]].


\bibitem{Hentschinski:2009zz} 
  M.~Hentschinski,
  arXiv:0908.2576 [hep-ph], DESY-THESIS-2009-025.
\bibitem{Hentschinski:2009ga} 
  M.~Hentschinski,
  Nucl.\ Phys.\ Proc.\ Suppl.\  {\bf 198}, 108 (2010)
  [arXiv:0910.2981 [hep-ph]].

\bibitem{Bartels:2004ef} 
  J.~Bartels, L.~N.~Lipatov and G.~P.~Vacca,
  Nucl.\ Phys.\ B {\bf 706}, 391 (2005)
  [hep-ph/0404110];
  G.~A.~Chirilli, L.~Szymanowski and S.~Wallon,
  Phys.\ Rev.\ D {\bf 83} (2011) 014020
  [arXiv:1010.0285 [hep-ph]].
\bibitem{Ayala:2014nza} 
  A.~Ayala, E.~R.~Cazaroto, L.~A.~Hern\'andez, J.~Jalilian-Marian and M.~E.~Tejeda-Yeomans,
  Phys.\ Rev.\ D {\bf 90}, no. 7, 074037 (2014)
  [arXiv:1408.3080 [hep-ph]];
  J.~Jalilian-Marian,
  Phys.\ Rev.\ D {\bf 85} (2012) 014037
  [arXiv:1111.3936 [hep-ph]].

\bibitem{Hatta:2005rn} 
  Y.~Hatta, E.~Iancu, L.~McLerran, A.~Stasto and D.~N.~Triantafyllopoulos,
  Nucl.\ Phys.\ A {\bf 764}, 423 (2006)
  [hep-ph/0504182];
  Y.~Hatta,
  Nucl.\ Phys.\ A {\bf 781}, 104 (2007)
  [hep-ph/0607126].

\bibitem{Bondarenko:2018pvv}
  S.~Bondarenko and M.~A.~Zubkov,
  arXiv:1801.08066 [hep-ph];
  S.~Bondarenko, L.~Lipatov, S.~Pozdnyakov and A.~Prygarin,
  Eur.\ Phys.\ J.\ C {\bf 77} (2017) no.9,  630
  [arXiv:1708.05183 [hep-th]];
  Eur.\ Phys.\ J.\ C {\bf 77} (2017) no.8,  527
  [arXiv:1706.00278 [hep-ph]].


\bibitem{Caron-Huot:2013fea}
  S.~Caron-Huot,
  JHEP {\bf 1505} (2015) 093
  [arXiv:1309.6521 [hep-th]].







\bibitem{Hentschinski:2011xg} 
  M.~Hentschinski,
  Nucl.\ Phys.\ B {\bf 859}, 129 (2012)
  [arXiv:1112.4509 [hep-ph]].



\bibitem{Antonov:2004hh} 
  E.~N.~Antonov, L.~N.~Lipatov, E.~A.~Kuraev and I.~O.~Cherednikov,
  Nucl.\ Phys.\ B {\bf 721}, 111 (2005)
  [hep-ph/0411185].


\bibitem{Steinmann}
O.~Steinmann,
Helv.\ Physica\ Acta {\bf 33} (1960) 257 –298.

\bibitem{Ayala:2017rmh}
  A.~Ayala, M.~Hentschinski, J.~Jalilian-Marian and M.~E.~Tejeda-Yeomans,
  Nucl.\ Phys.\ B {\bf 920} (2017) 232
  [arXiv:1701.07143 [hep-ph]];
  Phys.\ Lett.\ B {\bf 761} (2016) 229
  [arXiv:1604.08526 [hep-ph]].



\bibitem{McLerran:1994vd} 
  L.~D.~McLerran and R.~Venugopalan,
  Phys.\ Rev.\ D {\bf 50}, 2225 (1994)
  [hep-ph/9402335];
  A.~J.~Baltz, F.~Gelis, L.~D.~McLerran and A.~Peshier,
  Nucl.\ Phys.\ A {\bf 695}, 395 (2001)
  [nucl-th/0101024];
  F.~Gelis and A.~Peshier,
  Nucl.\ Phys.\ A {\bf 697}, 879 (2002)
  [hep-ph/0107142];
  I.~I.~Balitsky and A.~V.~Belitsky,
  Nucl.\ Phys.\ B {\bf 629}, 290 (2002)
  [hep-ph/0110158].





\bibitem{Gituliar:2015agu}
  O.~Gituliar, M.~Hentschinski and K.~Kutak,
  JHEP {\bf 1601} (2016) 181
  [arXiv:1511.08439 [hep-ph]];
  M.~Hentschinski, A.~Kusina, K.~Kutak and M.~Serino,
  arXiv:1711.04587 [hep-ph].

\bibitem{nlocgc} 
R.~Boussarie, A.~V.~Grabovsky, D.~Y.~Ivanov, L.~Szymanowski and S.~Wallon,
  Phys.\ Rev.\ Lett.\  {\bf 119} (2017) no.7,  072002
  [arXiv:1612.08026 [hep-ph]];
R.~Boussarie, A.~V.~Grabovsky, L.~Szymanowski and S.~Wallon,
  JHEP {\bf 1611} (2016) 149
  [arXiv:1606.00419 [hep-ph]];
G.~Beuf,
  Phys.\ Rev.\ D {\bf 96}, no. 7, 074033 (2017)
 [arXiv:1708.06557 [hep-ph]];
B.~Duclou\'e, H.~H\"anninen, T.~Lappi and Y.~Zhu,
  Phys.\ Rev.\ D {\bf 96}, no. 9, 094017 (2017)
  [arXiv:1708.07328 [hep-ph]].
 G.~A.~Chirilli, B.~W.~Xiao and F.~Yuan,
  Phys.\ Rev.\ D {\bf 86} (2012) 054005
  [arXiv:1203.6139 [hep-ph]].




\bibitem{Hentschinski:2005er} 
M.~Hentschinski, H.~Weigert and A.~Schafer,
Phys.\ Rev.\ D {\bf 73}, 051501 (2006)
[hep-ph/0509272].



\end{thebibliography}
\end{document}